\documentclass[12pt,epsf,amssymb,ulem,qsymbols]{article}
\usepackage{tabularx}
\usepackage{array}
\usepackage{graphics}
\usepackage{graphicx}
\usepackage{psfrag}
\usepackage{epsfig}
\usepackage{amsmath}
\usepackage{amssymb}
\usepackage{ulem}
\usepackage{setspace}
\usepackage{rotating}
\usepackage{colortbl}
\usepackage{tabularx}
\usepackage{longtable}
\usepackage{multirow}

\usepackage[T1]{fontenc}
\usepackage{geometry}
\usepackage[english]{babel}
\usepackage[cp1250]{inputenc}
\usepackage{tensor}
\usepackage{amsfonts}
\usepackage{tikz}
\usepackage{upgreek}
 
\makeatletter

\usepackage{verbatim}

\setlongtables

\newcommand{\msbar}{{\overline{\rm MS}}}

\newcommand{\bea}{\begin{eqnarray}}
\newcommand{\eea}{\end{eqnarray}}
\newcommand{\beq}{\begin{equation}}
\newcommand{\eeq}{\end{equation}}
\newcommand{\ec}{\end{center}}
\newcommand{\bc}{\begin{center}}
\newcommand{\gev}{{\rm GeV}}
\newcommand{\mev}{{\rm MeV}}

\newcommand{\pdir}{p\kern -5.2pt\raise 0.2ex\hbox {/}}

\newcommand{\vdir}{v\kern -5.75pt\raise 0.15ex\hbox {/}}
\newcommand{\kdir}{k\kern -5.75pt\raise 0.15ex\hbox {/}}
\newcommand{\epsdir}{\epsilon\kern -5.0pt\raise 0.15ex\hbox {/}}
\newcommand{\bvdir}{\bar{v}\kern -5.75pt\raise 0.15ex\hbox {/}}
\newcommand{\Ddir}{D\kern -7.75pt\raise 0.20ex\hbox {/}}
\newcommand{\Adir}{A\kern -7.75pt\raise 0.20ex\hbox {/}}
\newcommand{\ldir}{l\kern -5.0pt\raise 0.2ex\hbox{/}}
\newcommand{\varepsdir}{\varepsilon\kern -5.5pt\raise 0.15ex\hbox{/}}

\newcommand{\nf}{{N_{\rm f}}}

\newcommand{\nn}{\nonumber}
\newcommand{\Li}{\mathrm{Li}}

\makeatother

\begin{document}
\thispagestyle{empty} 
\begin{flushleft}\
\end{flushleft}
\begin{flushright}
\begin{tabular}{l}
{\tt LPT 13-90}\\
\end{tabular}
\end{flushright}
\begin{center}
\vskip 1.8cm\par
{\par\centering \textbf{\LARGE  
\Large \bf Lattice QCD and QCD Sum Rule determination  }}\\
\vskip .4cm\par
{\par\centering \textbf{\LARGE  
\Large \bf of the decay constants of $\eta_c$, $J/\psi$ and $h_c$  states }}\\
\vskip 1.05cm\par
{\scalebox{.8}{\par\centering \large  
\sc Damir Be\v{c}irevi\'c$^a$, Goran Duplan\v ci\'c$^b$, Bruno Klajn$^b$,} 
\\
\scalebox{.8}{\par\centering \large  
\sc Bla\v{z}enka Meli\'c$^{b}$ and Francesco Sanfilippo$^{c}$}
{\par\centering \vskip 0.65 cm\par}
{\sl 
$^a$~Laboratoire de Physique Th\'eorique (B\^at.~210)~\footnote{Laboratoire de Physique Th\'eorique est une unit\'e mixte de recherche du CNRS, UMR 8627.}\\
Universit\'e Paris Sud, F-91405 Orsay-Cedex, France.}\\
{\par\centering \vskip 0.25 cm\par}
{\sl 
$^b$~Rudjer Bo\v{s}kovi\'c Institute, Theoretical Physics Division\\
P.O.Box 180, HR-10002 Zagreb, Croatia.}\\ 
{\par\centering \vskip 0.25 cm\par}
{\sl 
$^c$~School of Physics and Astronomy, University of Southampton, \\ 
Highfield, Southampton, SO171 BJ, UK.}\\ 
{\vskip 1.05cm\par}}
\end{center}

\vskip 0.55cm
\begin{abstract}
We compute the decay constants of the lowest $c\bar c$-states with quantum numbers $J^{PC}=0^{-+}$ ($\eta_c$), $1^{--}$ ($J/\psi$), and $1^{+-}$ ($h_c$) by using lattice QCD and QCD sum rules. 
We consider the coupling of $J/\psi$ to both the vector and tensor currents. 
Lattice QCD results are obtained from the unquenched ($\nf=2$) simulations using twisted mass QCD at four lattice spacings, allowing us to take the continuum limit. 
On the QCD sum rule side we use the moment sum rules. 
The results are then used to discuss the rate of $\eta_c\to \gamma\gamma$ decay, and to comment on the factorization in $B\to X_{c\bar c} K$ decays, with  $X_{c\bar c}$ being either $\eta_c$ or $J/\psi$.
\end{abstract}
\vskip 2.6cm
{\small PACS: 12.38.Gc,14.40.Pq, 13.25.Gv, 11.55.Hx} 
\newpage
\setcounter{page}{1}
\setcounter{footnote}{0}
\setcounter{equation}{0}
\noindent

\renewcommand{\thefootnote}{\arabic{footnote}}

\setcounter{footnote}{0}
\section{\label{sec-0}Introduction}
Charmonium systems provide us with a playground for understanding the features of quark confinement,  for testing the validity of various quark models, and for describing processes that are interesting for the weak interaction phenomenology as well as for the search of physics beyond the Standard Model (BSM)~\cite{voloshin}.
Most of the quark models aim at describing the spectrum of charmonium states, including the orbital and radial excitations. Not many of these models, however, are used to describe the hadronic matrix elements as that requires a more detailed knowledge about the non-perturbative QCD dynamics of hadronic confinement. 

The method of QCD sum rules (QCDSR) was first tested on charmonium systems, and the fact that a number of $J^{PC}=1^{--}$ charmonium excitations have been detected and their electronic widths measured, was actually used to fix some of the QCDSR  parameters  relevant to the non-perturbative QCD effects expressed in terms of power corrections (QCD vacuum condensates)~\cite{Shifman:1978bx,Reinders:1981si,reviews}. In this paper we report on our results concerning the simplest matrix elements related to three charmonium states ($\eta_c$, $J/\psi$, $h_c$) and focus on four decay constants ($f_{\eta_c}$, $f_{J/\psi}$,  $f_{J/\psi}^T$, $f_{h_c}$) defined via,
\bea \label{def1}
&&\langle 0\vert \bar c (0) \gamma_\mu \gamma_5 c(0)
\vert \eta_c(p) \rangle = -i f_{\eta_c} p_\mu \;, \nn \\
&& \nn \\
&&\langle 0\vert \bar c (0) \gamma_\mu c(0)
\vert J/\psi (p,\lambda) \rangle = f_{ J/\psi} m_{ J/\psi} e_\mu^\lambda \,, \nn \\
&& \nn \\
&&\langle 0\vert \bar c (0) \sigma_{\mu\nu} c(0)
\vert J/\psi (p,\lambda) \rangle = i f_{ J/\psi}^T(\mu) \left(  e_\mu^\lambda p_\nu - e_\nu^\lambda p_\mu\right) \,, \nn\\
&&\nn  \\
&&\langle 0\vert \bar c (0) \sigma_{\mu\nu} c(0)
\vert h_c (p,\lambda) \rangle = i f_{ h_c}(\mu) \varepsilon_{\mu\nu\alpha\beta} e^\alpha_\lambda p^\beta\,,
\eea
where the $\mu$-dependence of the couplings to the tensor current indicates the renormalization scale and scheme dependence.

Of the above couplings only $f_{J/\psi}$ can be directly extracted from experiment via
\bea
\Gamma(J/\psi \to e^+e^-) = {4\pi\alpha_{\rm em}\over 3 m_{J/\psi} }\  {4\over 9} f_{J/\psi}^2.
\eea
The other couplings are not as directly related to experiment but they are still very relevant for phenomenology. 
For example,  $f_{\eta_c}$ enters decisively in the theoretical description of  the $\gamma^\ast \gamma^\ast \to \eta_c$ decay form factor, and of $\Gamma(\eta_c \to \gamma\gamma)$ in particular~\cite{models, kroll, lansberg, swanson1,melikhov}. 
Similarly, the phenomenological studies of the small-$x$ gluon distribution function from the inclusive production of $\eta_c$ requires the knowledge of $f_{\eta_c}$~\cite{1211}.  
Furthermore, such couplings can be helpful in describing the non-leptonic $B$-decays and to check for deviations between the measured and the results obtained by using the factorization approximation. 
For example, by combining the following decay modes~\cite{PDG}, 
\begin{align}\label{eq:00}
&B(B^+\to \eta_c K^+) = (9.6\pm 1.1)\times 10^{-4},&& B(B^0\to \eta_c K^0) = (7.9\pm 1.2)\times 10^{-4},  \nn\\
&B(B^+\to J/\psi K^+) = (1.03\pm 0.03)\times 10^{-3},&& B(B^0\to  J/\psi K^0) = (8.7\pm 0.3)\times 10^{-4}, 
\end{align}
one can use $f_{\eta_c}/f_{J/\psi}$ and the known information about the $B\to K$ form factors to check for validity of the factorization approximation. Otherwise, by imposing the factorization, one can get a useful information about the ratio of $B\to K$ form factors. 
On the other hand, a measurement of a non-zero $B(B^+\to h_c K^+)$, which is currently only bounded from above ($B(B^+\to h_c K^+)< 3.8\times 10^{-6}$~\cite{PDG}), could be interpreted as either a measurement of the deviation with respect to the factorization approximation, or a signal of the presence of coupling to the tensor operator that might appear only in the case of physics BSM.
Finally, the coupling $f_{J/\psi}^T$ may be interesting when checking for the presence of the New Physics operators in various processes.

In the first part of this paper we will discuss the computation of $f_{\eta_c}$, $f_{J/\psi}$,  $f_{J/\psi}^T$, $f_{h_c}$ by using the QCDSR. As we shall see the approximation of  `{\sl one resonance plus continuum}',  that we employ on the phenomenological side of sum rules, results in sizable error bars on the decay constants. In the second part, we compute the same quantities by means of numerical simulations of QCD on the lattice. Finally we compare our results and make a brief discussion of the impact of our results on two topics in phenomenology.

\section{\label{sec-1}Two-point QCD sum rules}
To estimate the hadronic properties of charmonium systems  (masses and decay constants) by means of QCDSR one needs to compute the two-point correlation functions. 
Here we will focus to the following three: 
\bea\label{eq:corr0}
&&\Pi_{\mu \nu} (q) = i \int dx\ e^{iqx}\langle 0\vert {\cal T}\left[ V_\mu^\dagger(x)V_\nu(0)\right]\vert 0\rangle \,,\nn\\
&&\Pi_{P} (q^2) = i \int dx\ e^{iqx}\langle 0\vert {\cal T}\left[ P^\dagger(x) P(0)\right]\vert 0\rangle \,,\nn\\
&&\Pi_{\mu \nu\rho\sigma} (q) = i \int dx\ e^{iqx}\langle 0\vert {\cal T}\left[ T_{\mu\nu}^\dagger(x)T_{\rho\sigma}(0)\right]\vert 0\rangle \,,
\eea
where $V_\mu =\bar c\gamma_\mu c$, $P = 2 m_c\  i\bar c\gamma_5 c$, and $T_{\mu\nu}=\bar c\sigma_{\mu\nu}c$, with $\sigma_{\mu\nu}=i/2\times [\gamma_\mu,\gamma_\nu]$.  
In terms of the Lorentz scalars the vector and tensor correlation functions can be written as:
\bea
&&\Pi_{\mu \nu} (q) = \left( {q_\mu q_\nu}-g_{\mu\nu} q^2\right) \Pi_V(q^2)\,,\nn\\
&& \Pi_{\mu \nu\rho\sigma}  (q) = P^-_{\mu \nu\rho\sigma} \Pi_-(q^2)+ P^+_{\mu \nu\rho\sigma}\Pi_+(q^2)\,,
\eea
where the projectors 
\bea\label{eq:projectors}
&&P^-_{\mu \nu\rho\sigma}  =  g_{\mu\sigma} {q_\nu q_\rho}+ g_{\nu\rho} {q_\mu q_\sigma} - g_{\mu\rho} {q_\nu q_\sigma}- g_{\nu\sigma} {q_\mu q_\rho} 
 \,,\nn\\
&& P^+_{\mu \nu\rho\sigma}  =q^2 \left ( g_{\mu\rho}g_{\nu\sigma} - g_{\mu\sigma} g_{\nu\rho} \right ) -  P^-_{\mu \nu\rho\sigma} \,,
\eea
separate the even and odd parity parts, and therefore $\Pi_+(q^2)$ will be used to discuss the $h_c(1^{+-})$ channel, while $\Pi_-(q^2)$ the ordinary $J/\psi(1^{--})$ state. 
Note that $P^i P^j =12 q^4\delta^{ij}$ in $d=4$ dimensions. For the perturbative part, each of the invariant functions $\Pi_i(q^2)$ ($i=P,V,+,-$) satisfies the dispersion relation, 
\bea\label{eq:dr}
\Pi_i(q^2)=\frac{1}{\pi}\int_0^\infty {{\rm Im}\Pi_i(s)\over s-q^2} ds \equiv \int_0^\infty {\rho_i(s)\over s-q^2} ds\,,
\label{eq:Pi}
\eea
with a suitable number of subtractions. Each spectral function, $\rho_i(s)$, is then computed in perturbation theory and can be written as,  
\bea
\rho_i^{\rm pert}(s) = \rho_i^{(0)}(s)+ {\alpha_s\over \pi}\rho_i^{(1)}(s)\,,
\eea
where the scale dependence is kept implicit. Besides the perturbative contribution, to the above $\Pi_i(q^2)$ one also needs to add the non-perturbative terms.  
The leading non-perturbative contributions to the correlation functions involving charmonia are power corrections proportional to the gluon condensate, $\langle \frac{\alpha_s}{\pi} G_{\mu \nu}^a G^{\mu\nu \; a} \rangle \equiv \langle \frac{\alpha_s}{\pi} G^2 \rangle $, namely, 
\bea
\Pi_i^{\rm non-pert}(q^2) =\left. C_i^{\rm G}(Q^2) \langle \frac{\alpha_s}{\pi} G^2 \rangle  \right|_{Q^2=-q^2},
\eea  
where the Wilson coefficients $C_i^{\rm G}(Q^2)\propto 1/Q^{2 n_i}$ are also computable perturbatively, with $n_i>0$ depending on the operators used.  

Complete expressions for the spectral functions $\rho_{i}^{pert}(s)$ as well as for gluon condensate contributions $C_i^{\rm G}(Q^2)$ are 
collected in Appendix, where a brief discussion about the calculation can be found.~\footnote{A more detailed description of the calculation, as well as the expressions for other charmonium states, will be given in a separate publication.}

When studying charmonia it is convenient to use the so-called moment sum rules~\cite{Shifman:1978bx,Reinders:1981si, reviews}. One starts by defining the moments of eq.~(\ref{eq:dr}),
\bea\label{eq:mom1}
{\cal M}_n(Q_0^2) =  \left.{1\over n!}\left({d\over dq^2}\right)^n \Pi_i(q^2)\right|_{q^2=-Q^2_0}= 
\int_{4 m_c^2}^\infty {\rho_i^{\rm pert}(s)\over (s+Q_0^2)^{n+1}} ds\,,
\eea
at some spacelike $Q_0^2$, far from the resonance region. In practice $Q_0^2$ is a parameter that is to be adjusted in order to improve the convergence of the integral on the right hand side (r.h.s.) of the above equation. Since the mass of the charm quark is large with respect to  $\Lambda_{\rm QCD}$ it is customary to use $Q_0^2=4 m_c^2 \xi$, and by changing the integration variable in the dispersion relation, $s\to v^2=1- 4m_c^2/s$, the theory part of the $n^{th}$ moment can be written as,  
\begin{align}\label{eq:mom2}
&{\cal M}^{{\rm theo.}\ i}_n(\xi) = {\cal M}_n^{\rm pert.}(\xi) + {\cal M}_n^{\rm non-pert.}(\xi) \nn \\
&\qquad =
\frac{1}{(4m_c^2)^n}\int_{0}^1 {2 v (1-v^2)^{n-1} \rho_i(v) \over \left[ 1 + \xi (1-v^2) \right]^{n+1}  }\ dv +\left. \frac{1}{n!} \left( - \frac{d}{dQ^2} \right)^n    C_i^{\rm G}(Q^2)\ \langle \frac{\alpha_s}{\pi} G^2 \rangle \right|_{Q^2=Q_0^2=4 m_c^2 \xi}.
\end{align}
On the other hand, the same moments (\ref{eq:mom1}) can be expressed in terms of hadronic quantities. By inserting all possible hadronic states $H$ in the correlators~(\ref{eq:corr0}) that can couple to each of the above operators, one can write
\bea\label{eq:mom3}
{\cal M}_n^{{\rm phen.}\ i}(Q_0^2) =  \sum_{k=0}^\infty \frac{ | \langle 0 | J^i(0) | H_k\rangle |^2 }{  \left (m_{H_k}^2 + Q_0^2 \right )^{n+1}} \,,
\eea  
where the sum runs over all possible single or multiparticle hadronic states, and $J^i$ stands for a generic bilinear quark operator. The situations in which the masses and couplings of the higher excited states in the sum~(\ref{eq:mom3}) are experimentally established are extremely rare. A notable example is that of the first few $J^{PC}=1^{--}$ states for which both the masses and electronic widths, $\Gamma(\psi(nS)\to e^+e^-)$, have been measured. This information was used to fix the value of the gluon condensate in ref.~\cite{Shifman:1978bx}, and then further refined in ref.~\cite{Ioffe:2002be}.~\footnote{For a recent review concerning the various estimates of the gluon condensate, please see ref.~\cite{narison}.} In the most phenomenologically relevant situations, however, only the position of the first pole in the sum~(\ref{eq:mom3}) is known, whereas for the rest of the sum one invokes the quark-hadron duality and replaces them by the spectral function $\rho_i^{\rm pert}(s)$ in the dispersion relation, starting from some threshold, $s_0^i > m_{H_0^{(i)}}^2$. After using the definitions~(\ref{def1}), we have
\bea\label{eq:ImPi_hadr}
{\cal M}_n^{{\rm phen.}\ V}(Q_0^2) = \frac{f_{J/\psi}^2}{\left (m_{J/\psi}^2 + Q_0^2 \right )^{n+1}} + \int_{s_0^{\psi}}^{\infty} \frac{\rho_V^{\rm pert.}(s) ds}{\left (s + Q_0^2 \right )^{n+1}} \,,\nonumber \\
{\cal M}_n^{{\rm phen.}\ P}(Q_0^2) = \frac{ \left (f_{\eta_c}  m_{\eta_c}^2\right)^2}{\left (m_{\eta_c}^2 + Q_0^2 \right )^{n+1}} 
+ 4 m_{c}^2 \int_{s_0^{\eta_c}}^{\infty} \frac{\rho_P^{\rm pert.}(s) ds}{\left (s + Q_0^2 \right )^{n+1}}\nn \,,\\
{\cal M}_n^{{\rm phen.}\ +}(Q_0^2) =  \frac{f_{h_c}^2}{\left (m_{h_c}^2 + Q_0^2 \right )^{n+1}} +
\int_{s_0^{h_c}}^{\infty} \frac{\rho_+^{\rm pert.}(s) ds}{\left (s + Q_0^2 \right )^{n+1}}\nn \,,\\
{\cal M}_n^{{\rm phen.}\  -}(Q_0^2) =  \frac{[f_{J/\psi}^{T}(\mu)]^2}{\left (m_{J/\psi}^2 + Q_0^2 \right )^{n+1}} +
\int_{s_0^{\psi^T}}^{\infty} \frac{\rho_-^{\rm pert.}(s) ds}{\left (s + Q_0^2 \right )^{n+1}}\,,
\eea
where the renormalization scale is chosen to be $\mu^2= m_c^2 + Q_0^2$, with $m_c\equiv m_c^\msbar(m_c)$. After equating eqs.~(\ref{eq:mom2}) and (\ref{eq:ImPi_hadr}) we can define
\begin{align}
\widetilde {\cal M}_n^{i}(\xi,s_0) =  \frac{1}{(4m_c^2)^n}\int_{0}^{v[s_0^i]} {2 v (1-v^2)^{n-1} \rho_i^{\rm pert.}(v) \over \left[ 1 + \xi (1-v^2) \right]^{n+1}  }\ dv  +\left. \frac{1}{n!} \left( - \frac{d}{dQ^2} \right)^n    C_i^{\rm G}(Q^2)\ \langle \frac{\alpha_s}{\pi} G^2 \rangle \right|_{Q^2=4 m_c^2 \xi},
\end{align}
where $v[s_0]= \sqrt{1- 4m_c^2/s_0}$, so that 
\begin{align}
m_{J/\psi}^2 = - 4m_c^2\xi +{  \widetilde {\cal M}_n^{V}(\xi,s_0^\psi)  \over  \widetilde {\cal M}_{n+1}^{V}(\xi,s_0^\psi) },\qquad 
f_{J/\psi} = \left (m_{J/\psi}^2 + 4 m_c^2 \xi \right )^\frac{n+1}{2} \left[ \widetilde {\cal M}_n^{V}(\xi,s_0^\psi) \right]^{1/2}\,,\nn
\end{align}
\begin{align}
m_{\eta_c}^2 = - 4m_c^2\xi +{  \widetilde {\cal M}_n^{P}(\xi,s_0^{\eta_c})  \over  \widetilde {\cal M}_{n+1}^{P}(\xi,s_0^{\eta_c}) },\qquad 
f_{\eta_c} = \left (m_{\eta_c}^2 + 4 m_c^2 \xi \right )^\frac{n+1}{2} \left[ \widetilde {\cal M}_n^{P}(\xi,s_0^{\eta_c}) \right]^{1/2}\frac{2 m_c}{m_{\eta_c}^2},\nn
\end{align}
\begin{align}
m_{h_c}^2 = - 4m_c^2\xi +{  \widetilde {\cal M}_n^{+}(\xi,s_0^{h_c})  \over  \widetilde {\cal M}_{n+1}^{+}(\xi,s_0^{h_c}) },\qquad 
f_{h_c}(\mu_0) =  \left.\left (m_{h_c}^2 + 4 m_c^2 \xi \right )^\frac{n+1}{2} \left[ \widetilde {\cal M}_n^{+}(\xi,s_0^{h_c}) \right]^{1/2} \right|_{\mu_0=m_c\sqrt{1+4\xi} }\,,\nn
\end{align}
\begin{align}\label{eq:SRfinal}
m_{J/\psi}^2 = - 4m_c^2\xi +{  \widetilde {\cal M}_n^{-}(\xi,s_0^\psi)  \over  \widetilde {\cal M}_{n+1}^{-}(\xi,s_0^\psi) },\qquad 
f_{J/\psi}^T(\mu_0) = \left. \left (m_{J/\psi}^2 + 4 m_c^2 \xi \right )^\frac{n+1}{2} \left[ \widetilde {\cal M}_n^{-}(\xi,s_0^\psi) \right]^{1/2} \right.\,.
\end{align}
In other words the masses are obtained from the ratios of moments, while the decay constants are computed from one or several moments separately. Before discussing the practical procedure we use to get the results for the decay constants we need to stress that: (1) ${\cal O}(\alpha_s)$ corrections to the functions $\Pi_\pm(Q_0^2)$ are new. In ref.~\cite{Reinders:1981si} the authors computed $\rho_+^{(1)}(s)$, by using  the operator with a single derivative, $\overline{c} \partial_{\mu} \gamma_5 c$, instead of the tensor density. Their $\rho_+^{(1)}(s)$ agrees with ours, apart from the correction coming from anomalous dimension of the tensor current.  Instead, $\rho_-^{(1)}(s)$ is completely new; (2) Our results for  $C_G^i(Q^2)$ agree with those presented in refs.~\cite{Reinders:1981si, radyushkin,broadhurst}. Here again, the result for $C_G^-(Q^2)$ is new.

\subsection{\label{sec-1bis}Evaluation of Sum Rules}

In this section we discuss the evaluation of  QCDSRs given in eq.~(\ref{eq:SRfinal}). Our strategy for all sum rules, except the one for $h_c$, consists in requiring that the mass of the lowest lying hadron obtained from the ratios of moments coincides with the experimentally established value to less then or equal to $1\%$. Only in the case of the sum rule for $h_c$ we allow that agreement to be within $5\%$. For the reader's convenience we quote the masses of the three lowest lying states we discuss in this paper~\cite{PDG}:
\bea
m_{\eta_c}^{\rm exp.}=2.984 \ \gev,\quad 
m_{J/\psi}^{\rm exp.}=3.0969\ \gev,\quad
m_{h_c}^{\rm exp.}=3.525\ \gev\,.
\eea
In evaluating the left hand side (l.h.s.) of eq.~(\ref{eq:SRfinal}) we take into account  the charm quark mass and the value of the gluon condensate from ref.~\cite{Ioffe:2002be},
\bea\label{eq:ioffe}
m_c^\msbar(m_c)=1.275(15)\ \gev,\qquad  \langle \frac{\alpha_s}{\pi} G^2 \rangle =0.009(7)\ \gev^4\,,
\eea
that are found to be highly correlated  (cf. fig.~5 in ref.~\cite{Ioffe:2002be}).~\footnote{This value of  the charm quark mass is consistent with the recent estimates of ref.~\cite{Dehnadi:2011gc}.
}
  We take that correlation into account and also vary the threshold parameter $s_0$ above the square of the  mass of the lowest state and its first radial excitation. More specifically, 
  \bea
  s_0^{\eta_c} \in [3.1^2, 3.5^2]\ \gev^2,\quad  s_0^\psi \in [3.3^2, 3.65^2]\ \gev^2,\quad   s_0^{h_c} \in [3.6^2, 4.0^2]\ \gev^2.
  \eea
While $m_{\eta_c^\prime}^{\rm exp.}=3.639(1)$~GeV and $m_{\psi^\prime}^{\rm exp.}=3.686$~GeV are known~\cite{PDG}, the first radial excitation of the $h_c$ state could be extracted from lattice QCD study of ref.~\cite{Liu:2012ze}, $m_{h_c^\prime}^{\rm latt.}=3.639(1)$~GeV. With the sum rule parameters [$m_c^\msbar(m_c)$, $\langle \frac{\alpha_s}{\pi} G^2 \rangle $, $s_0^i$] varied in the intervals indicated above, we then look for the moments $n$ such that $\delta m^{\rm QCDSR}_{\eta_c,J/\psi}/m^{\rm exp.}_{\eta_c,J/\psi} \leq 1 \%$ and $\delta m^{\rm QCDSR}_{h_c}/m^{\rm exp.}_{h_c} \leq 5 \%$. 

  Furthermore we impose the standard QCDSR requirements, namely that the next-to-leading order correction to the moments represents less than $30\%$ with respect to the leading order term, and that the contribution coming from the gluon condensate does not exceed $50\%$ of the perturbative part. For the former requirement it is important to work with $\xi\neq 0$. We actually checked that for two values, $\xi=1$ and $\xi=2$, the range of values for the moments is such that the above criteria are fulfilled and the resulting values for the decay constants remain unchanged. The only exception is the sum rule for $h_c$, for which the mass of $h_c$, as obtained from the ratios of moments,  is always larger than the physical one. This excess is less than $5\%$ only for lower moments and for lower values of the threshold parameter $s_0$. For that reason in the discussion of our results for $m_{h_c}$ and $f_{h_c}$ we will vary $s_0^{h_c} \in [3.6^2, 3.8^2]\ \gev^2$. We attempted to enlarge the window in $s_0$, but the impact on our final results was only marginal.  
We should stress that the value of $n$ is not fixed to be common to all $s_0$, but they were found for each $s_0$ separately. Therefore this somewhat implicitly corresponds to a strategy adopted in ref.~\cite{dima2} in which the threshold parameter was considered to be a function of the Borel parameter (or equivalently of $n$ in the case of the moment sum rules). We do not introduce any extra parameter but verify that $s_0$ and $n$ are indeed strongly correlated. 
  
\begin{figure}[t!]
\begin{center}
\begin{tabular}{@{\hspace{-0.85cm}}cc}
{\resizebox{8cm}{!}{\includegraphics{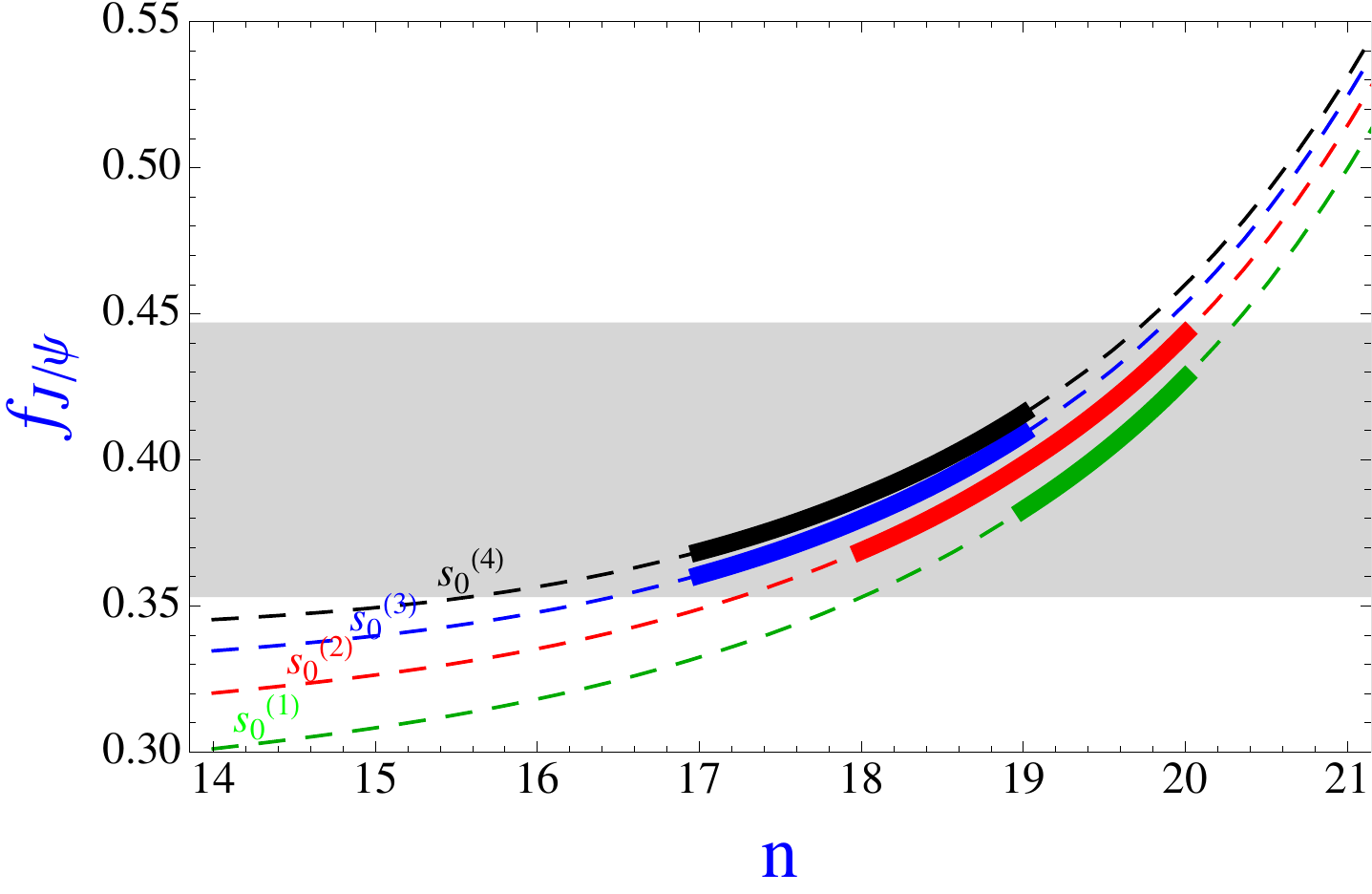}}}&{\resizebox{8cm}{!}{\includegraphics{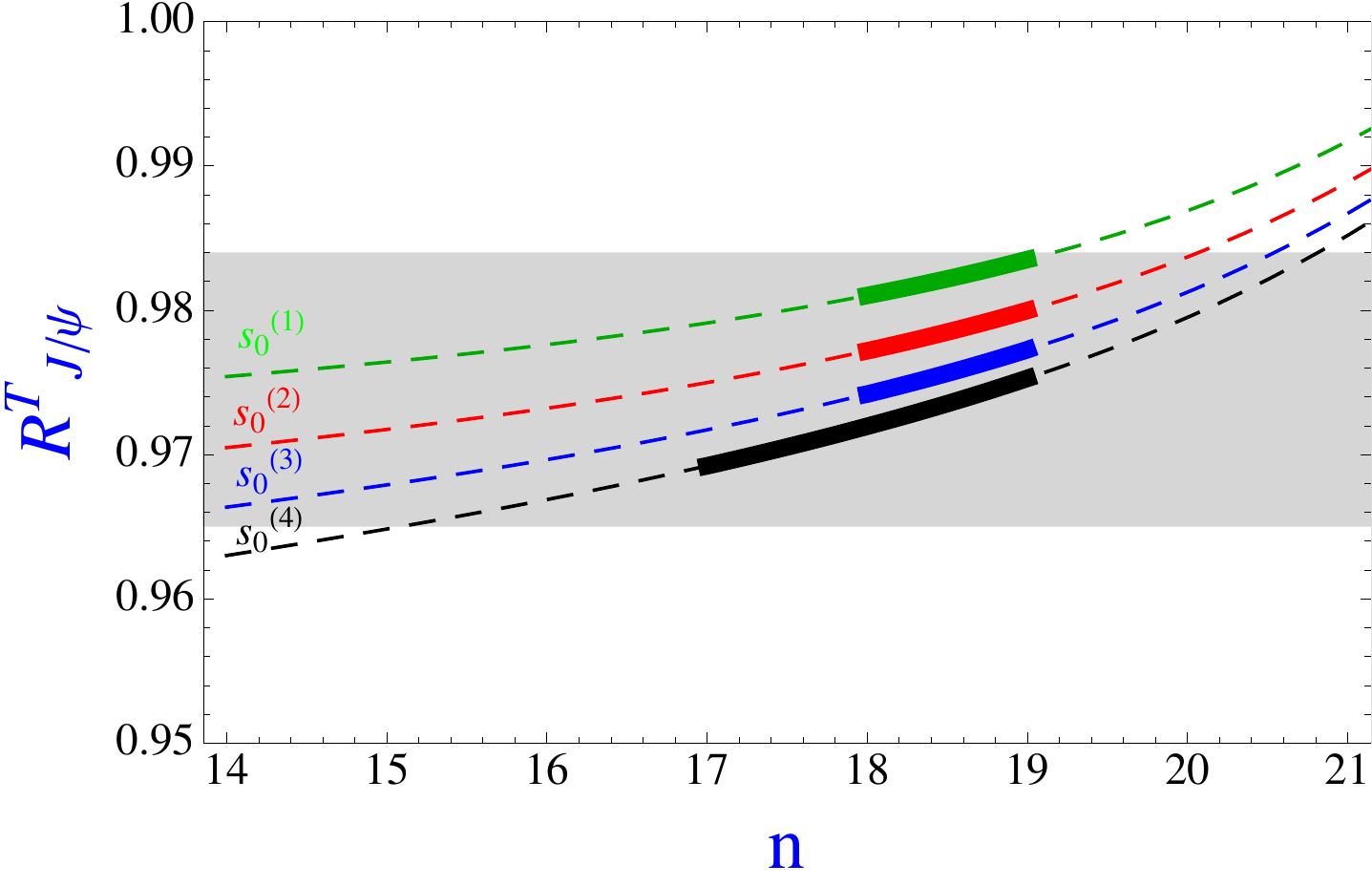}}} \\
{\resizebox{8cm}{!}{\includegraphics{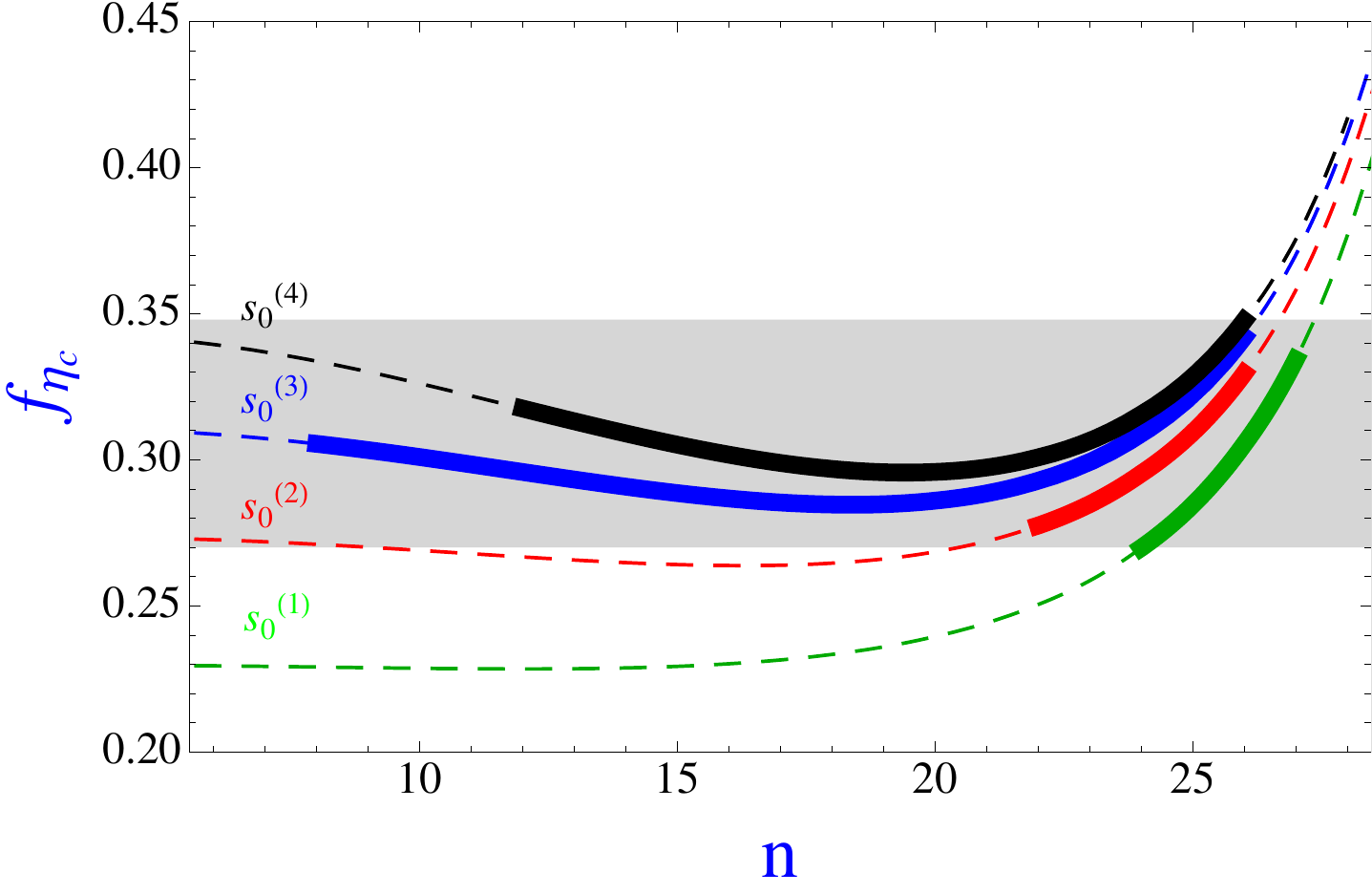}}}&{\resizebox{8cm}{!}{\includegraphics{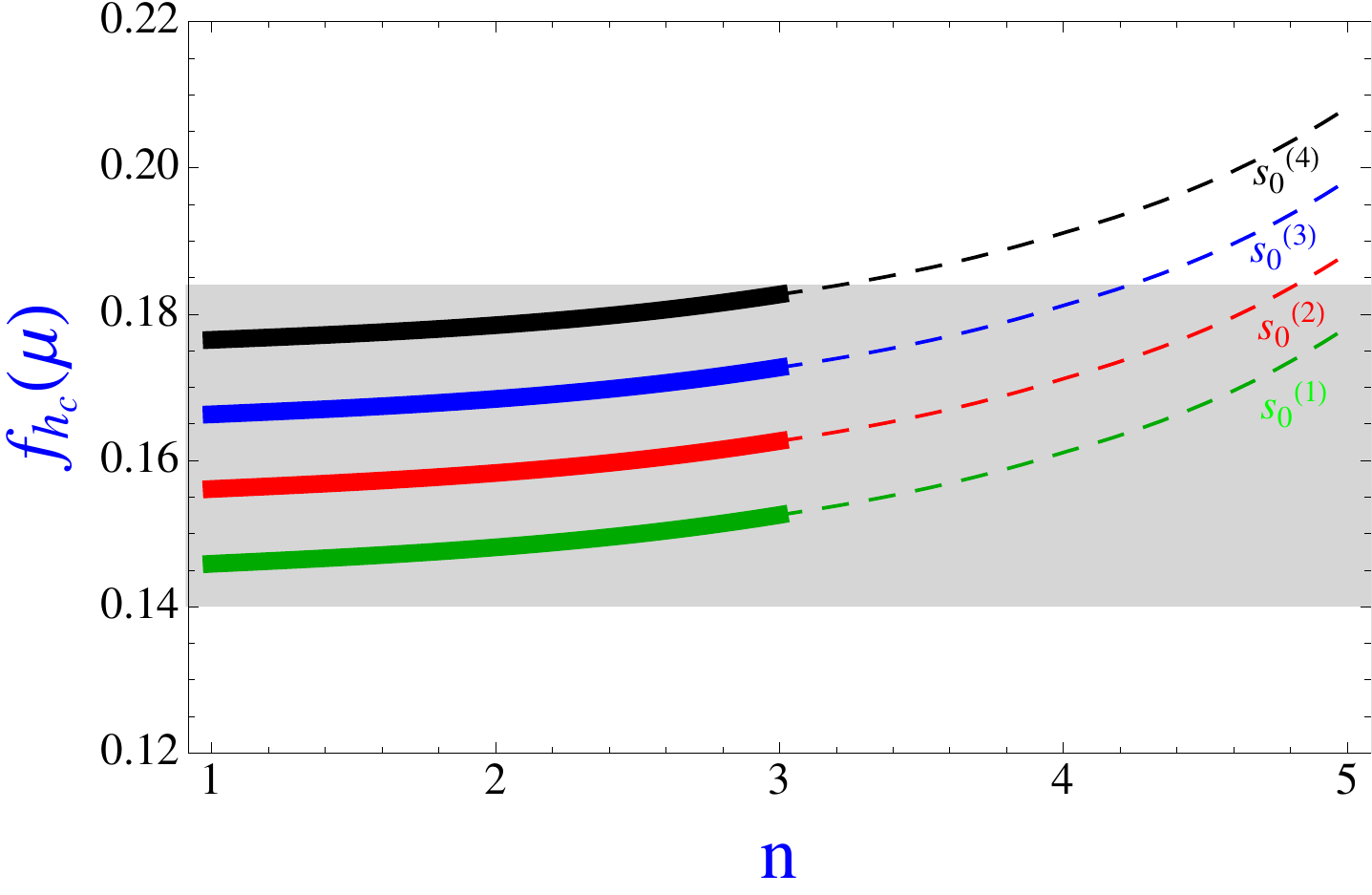}}} \\
\end{tabular}
\vspace*{-.1cm}
\caption{\label{fig:3}{\footnotesize 
Couplings $f_{\eta_c}$, $f_{h_c}(2\ \gev)$, $f_{J/\psi}$ (in GeV), and the ratio $R_{J/\psi}^T(2\ \gev)$ computed by means of the moment sum rules. Thick lines correspond to the moments satisfying the requirement that 
$\delta m^{\rm QCDSR}_{\eta_c,J/\psi}/m^{\rm phys.}_{\eta_c,J/\psi}\leq 1\%$, and $\delta m^{\rm QCDSR}_{h_c}/m^{\rm phys.}_{h_c}\leq 5\%$.  Illustration is provided for the central values of the charm quark mass and the gluon condensate, and for four equidistant values of the threshold parameter $s_0\in [s_0^{(1)},s_0^{(4)}]$. Shaded area display the range of values obtained after varying all the QCDSR parameters. 
 } }
\end{center}
\end{figure}

For example, and by using the central value of the charm quark mass and of the gluon condensate~(\ref{eq:ioffe}), and by varying $s_0^i$, we find that all the above criteria are satisfied for
 \bea
\eta_c:  n\in [12,26]\,,\qquad J/\psi :   n\in [17,19]\,,\qquad h_c : n\in [1,3] \,.
 \eea
The above ranges of $n$ are found for $\xi=2$. They change with the value of $\xi$ and for larger $\xi$ the values of $n$ satisfying our criteria become larger. With these values of $n$ we then compute the decay constants. Illustration of the stability of the sum rule results is provided in fig.~\ref{fig:3}. 
We should note that each decay constant is highly sensitive to the mass of the hadron. To make the procedure fully self-consistent, in the evaluation of the sum rule for each decay constant we use the corresponding hadron mass obtained by the same sum rule. Had we used the physical mass of the hadron state instead of the one obtained from the sum rule, the resulting curves  in fig.~\ref{fig:3} would be considerably flatter. 
It turns out that the variation of the threshold parameter $s_0$ already covers most of the allowed values for the decay constants that are shown by the shaded areas in fig.~\ref{fig:3}.  
Note that these shaded intervals in fig.~\ref{fig:3} are obtained by varying all of the QCD sum rule parameters: $s_0$, $n$, $m_c^\msbar(m_c)$, $\langle \frac{\alpha_s}{\pi} G^2 \rangle$, and for $\xi \in\{1,2\}$.  
Another important comment is that we take into account the correlation between the charm quark mass and of the gluon condensate found in ref.~\cite{Ioffe:2002be}. In that latter paper the values of the charm quark mass and of the gluon condensate have been obtained from the vector-vector sum rule by using the three-loop perturbative expressions, and by including the loop corrections to the Wilson coefficient that multiplies the gluon condensate. More importantly, as far as the stability of the results is concerned, a rich experimental information about the spectral function in the ${\cal M}_n^{{\rm phen.}\ V}(Q_0^2)$ has been included. 
We do not aim at that level of accuracy. Instead we content ourselves by working with the two-loop QCD expressions on the perturbative side, and only one resonance has been included to the hadronic side of the sum rules before evoking the quark-hadron duality. For that reason the expected accuracy of the sum rules on the decay constant will be relatively modest. Indeed we get
\bea\label{eq:QCDSRfV}
f_{J/\psi} = (335 \div 447)\ \mev = (401 \pm 46)\ \mev \,, 
\eea
therefore with about $10\%$ uncertainty, which is a typical accuracy of the sum rule computation of the hadronic  decay constants~\cite{reviews}. Note again that $335$~MeV and $447$~MeV correspond to the minimal and maximal value of  $f_{J/\psi}$ obtained from the QCD sum rule after varying {\it all the parameters} in a way described above. 

As for $f_{\eta_c}$ we find 
\bea\label{eq:QCDSRfP}
f_{\eta_c} = (270 \div 348)\ \mev = (309 \pm 39)\ \mev \,, 
\eea
which is somewhat lower than $f_{\eta_c}= 356(16)$~MeV found in ref.~\cite{DiSalvo:1994dg} in which the authors made additional assumptions about the contributions  to the phenomenological side of the spectral function coming from the radially excited $\eta_c$, or $f_{\eta_c}= 346(33)$~MeV  found in ref.~\cite{,Braguta:2006wr} where the threshold parameter was assumed to be much larger than the lowest radially excited states.~\footnote{Our $f_{\eta_c}$ is related to $g_1$ from ref.~\cite{DiSalvo:1994dg} as $f_{\eta_c} = \frac{3}{2} m_{\eta_c}/g_1$. } In the earlier sum rule estimates another definition has been used, related to ours via $g_{\eta_c}= f_{\eta_c}/2 m_c$~\cite{novikov}. That definition is renormalization scale dependent but since the authors of ref.~\cite{novikov} used only the leading order expressions for the perturbative part of the spectral function, the choice of the scale and scheme could not be specified.  Note also that in the past  the computations were often done by using the pole charm quark mass, so that the approximation $2 M_c\approx m_{\eta_c}$ was justified. In that way the resulting value for $f_{\eta_c}$ was larger.  Finally, the recent computation of this decay constant by using the Borel sum rule and somewhat different criteria for the choice of the sum rule parameters, lead to a much larger value~\cite{recent-2}.

Concerning the coupling  $f_{J/\psi}^T(\mu)$, the discussion of this sum rule is qualitatively and quantitatively very similar to that of  $f_{J/\psi}$ obtained from the vector-vector correlation function, and the moments for which the criteria discussed above are satisfied is essentially the same. We obtain,
\bea\label{eq:QCDSRfT}
f_{J/\psi}^T(2\ \gev) = (346 \div 436)\ \mev = (391 \pm 45)\ \mev \,, 
\eea
in the $\msbar$ scheme. Our result agrees with $f_{J/\psi}^T = (408 \pm 26)$~MeV, presented in ref.~\cite{Braguta:2007fh}. In contrast to the latter paper we could also specify the renormalization scale at which $f_{J/\psi}^T(\mu)$ is defined because we included in our calculation the next-to-leading order QCD correction to the spectral function (cf. discussion in the appendix of the present paper). Since the behavior of $f_{J/\psi}^T(\mu)$ with respect to the variation of the QCD sum rule parameters is very similar to that of $f_{J/\psi}$, it is more judicious to compute the ratio of the two, 
\bea
R_{J/\psi}^T={f_{J/\psi}^T(\mu)\over f_{J/\psi}}\,,
\eea
which is illustrated in fig.~\ref{fig:3} where we see that the ratio $R_{J/\psi}^T$ is much more accurately estimated. By following the same criteria discussed above, and by using $\mu = 2$~GeV, we get 
\bea
R_{J/\psi}^T= (0.965 \div 0.984)  = 0.975\pm 0.010\,.
\eea
We emphasize that this is the first determination of $f_{J/\psi}^T$ and  $R_{J/\psi}^T$.

Finally, our result for the decay constant of the $h_c$ state, is
\bea\label{eq:reshc}
f_{h_c}(2 \ \gev) =  (140 \div 184)\ \mev = (162 \pm 22)\ \mev \,. 
\eea
We stress that this result is obtained for low moments and that for larger moments the sum rules progressively deteriorates in the sense that the mass of the lowest lying state becomes much larger and the decay constant much smaller. To our knowledge, up to now, the only QCDSR analysis of the $h_c$ state has been made in ref.~\cite{Wang:2012gj} in which the author reported $f_{h_c}=490(60)$~MeV, in clear disagreement with our value~(\ref{eq:reshc}).  We compared our expressions with those presented in ref.~\cite{Wang:2012gj}, and realized that the author of  ref.~\cite{Wang:2012gj} has calculated only a part of $\Pi_+(q^2)$ obtained using only the first part of  $P^+_{\mu \nu \rho \sigma}$ in (\ref{eq:projectors}), and therefore his result does not correspond to any physical state.

\section{\label{sec-2}Lattice QCD results}
We now compute the same quantities discussed above but by means of numerical simulations of QCD on the lattice. To that end we use the gauge field configurations generated by European Twisted Mass Collaboration (ETMC), in which the effect of $\nf=2$ dynamical (``{\sl sea}") light quarks have been included by using the Wilson regularization of QCD on the lattice with the maximally twisted mass term, namely~\cite{fr}~\footnote{Note that  the action is written in the ``physical basis" and not in the twisted one. }
\bea\label{eq:tmqcdC}
S=a^4\sum_{x}\bar \psi(x)\left\{ \frac{1}{2}\sum_\mu\gamma_\mu\left(\nabla_\mu+\nabla_\mu^\ast\right)-i\gamma_5\tau^3 r \left[m_{\rm cr}-\frac{a}{2}\sum_\mu\nabla_\mu^\ast\nabla_\mu\right]+\mu_c\right\}\psi(x)\,,
\eea
where $\nabla_\mu$ ($\nabla_\mu^\ast$) stands for the forward (backward) covariant derivative,  $m_{\rm cr}$ is the critical mass term tuned to restore the chiral symmetry of the massless action, otherwise broken by the Wilson term (also in the brackets), and $\mu_c$ is the bare charm quark mass. In the above action $\psi(x) = [c(x)\ c^\prime(x)]^T$ is a doublet of the charm quark field and its replica. The factor $i \gamma_5\tau^3 r$ cures the pathology of the standard Wilson quark action by rotating the Wilson term to the imaginary axis which is why one can simulate with sea quark masses considerably closer to the chiral limit. The quark propagators $S_c(0,0;\vec x,t)$ and $S_c^\prime(0,0;\vec x,t)$  are then obtained by inverting the above Wilson-Dirac operator with $r$ and $-r$, respectively. In practice $r=1$. Finally, we should mention that the action~(\ref{eq:tmqcdC}) refers to the valence charm quarks, but the same one is used to generate the gauge field configurations but with $\mu_c\to \mu_{q}$, mass of the light sea quark.  Detailed information about the lattices used in this work are given in tab.~\ref{tab:1}.
\begin{table}[t!!]
\centering 
{\scalebox{.91}{\begin{tabular}{|c|cccccc|}  \hline \hline
{\phantom{\huge{l}}}\raisebox{-.2cm}{\phantom{\Huge{j}}}
$ \beta$& 3.8 &  3.9  &  3.9 & 4.05 & 4.2  & 4.2    \\ 
{\phantom{\huge{l}}}\raisebox{-.2cm}{\phantom{\Huge{j}}}
$ L^3 \times T $&  $24^3 \times 48$ & $24^3 \times 48$  & $32^3 \times 64$ & $32^3 \times 64$& $32^3 \times 64$  & $48^3 \times 96$  \\ 
{\phantom{\huge{l}}}\raisebox{-.2cm}{\phantom{\Huge{j}}}
$ \#\ {\rm meas.}$& 240 &  240  & 150  & 150 & 150 & 100  \\ \hline 
{\phantom{\huge{l}}}\raisebox{-.2cm}{\phantom{\Huge{j}}}
$\mu_{\rm sea 1}$& 0.0080 & 0.0040 & 0.0030 & 0.0030 & 0.0065 &  0.0020   \\ 
{\phantom{\huge{l}}}\raisebox{-.2cm}{\phantom{\Huge{j}}}
$\mu_{\rm sea 2}$& 0.0110 & 0.0064 & 0.0040 & 0.0060 &   &     \\ 
{\phantom{\huge{l}}}\raisebox{-.2cm}{\phantom{\Huge{j}}}
$\mu_{\rm sea 3}$&  & 0.0085 &  & 0.0080 &   &     \\ 
{\phantom{\huge{l}}}\raisebox{-.2cm}{\phantom{\Huge{j}}}
$\mu_{\rm sea 4}$&  & 0.0100 &  &   &   &     \\   \hline 
{\phantom{\huge{l}}}\raisebox{-.2cm}{\phantom{\Huge{j}}}
$a \ {\rm [fm]}$&   0.098(3) & 0.085(3) & 0.085(3) & 0.067(2) & 0.054(1) & 0.054(1)      \\ 
{\phantom{\huge{l}}}\raisebox{-.2cm}{\phantom{\Huge{j}}}
$Z_T^\msbar (g_0^2,2\ \gev)$~\cite{ZZZ}& 0.73(2) & 0.750(9) & 0.750(9)  & 0.798(7) & 0.822(4) & 0.822(4)  \\ 
{\phantom{\huge{l}}}\raisebox{-.2cm}{\phantom{\Huge{j}}}
$Z_A (g_0^2)$~\cite{ZZZ}& 0.746(11) & 0.746(6) & 0.746(6)  & 0.772(6) & 0.780(6) & 0.780(6) \\ 
{\phantom{\huge{l}}}\raisebox{-.2cm}{\phantom{\Huge{j}}}
$\mu_{c}$~\cite{Blossier:2010cr}& 0.2331(82)  &0.2150(75)  &0.2150(75)   & 0.1849(65) & 0.1566(55) & 0.1566(55)  \\ 
 \hline \hline
\end{tabular}}}
{\caption{\footnotesize  \label{tab:1} Summary of the lattice ensembles used in this work (more information can be found in ref.~\cite{boucaud}).  Data obtained at different $\beta$'s are rescaled by $r_0/a$, and the overall lattice spacing is fixed by matching $f_\pi$ computed on the lattice with its physical value, leading to  $r_0= 0.440(12)$~fm (c.f. ref.~\cite{Blossier:2010cr}). All quark masses are given in lattice units.}}
\end{table}

\begin{table}[t!!]
\centering 
{\scalebox{.9}{\begin{tabular}{|c|ccccccc|}  \hline 
{\phantom{\huge{l}}}\raisebox{-.3cm}{\phantom{\Huge{j}}}
$(\beta,\mu_{\rm sea},L)$ &  $a m_{\eta_c}$  & $R_{J/\psi}$ & $R_{h_c}$ & $ f_{\eta_c}$ & $f_{J/\psi}$  & $f_{J/\psi}^T(\mu)$  & $f_{h_c}^T(\mu)$    \\    \hline 	 \hline 	
{\phantom{\huge{l}}}\raisebox{-.2cm}{\phantom{\Huge{j}}}
(3.80, 0.0080, 24) & 1.2641(2) & 1.0749(6) & 1.254(5)  &  0.388(12)&  0.464(14)&0.444(14) & 0.216(7) \\	            
{\phantom{\huge{l}}}\raisebox{-.2cm}{\phantom{\Huge{j}}}                                      
(3.80, 0.0110, 24) & 1.2645(3) & 1.0749(4) & 1.265(5)  &  0.387(12)&  0.460(15) &0.442(15) & 0.224(7) \\	 \hline            
{\phantom{\huge{l}}}\raisebox{-.2cm}{\phantom{\Huge{j}}}
(3.90, 0.0040, 24) & 1.1308(4) & 1.0621(5) & 1.235(6)  &  0.378(10)&  0.435(12)&0.413(12) & 0.213(13) \\	            
{\phantom{\huge{l}}}\raisebox{-.2cm}{\phantom{\Huge{j}}}                                      
(3.90, 0.0064, 24) & 1.1311(2) & 1.0628(4) & 1.235(6)  &  0.381(10)&  0.440(12) &0.420(12) & 0.213(10) \\	            
{\phantom{\huge{l}}}\raisebox{-.2cm}{\phantom{\Huge{j}}}                                      
(3.90, 0.0085, 24) & 1.1317(3) & 1.0630(4) & 1.245(3)  &  0.383(10)& 0.444(12) &0.426(12 & 0.230(10)\\	            
{\phantom{\huge{l}}}\raisebox{-.2cm}{\phantom{\Huge{j}}}                                      
(3.90, 0.0100, 24) & 1.1310(3) & 1.0632(4) & 1.240(5)  &  0.380(10)& 0.438(12)& 0.413(11) & 0.209(8) \\	            
{\phantom{\huge{l}}}\raisebox{-.2cm}{\phantom{\Huge{j}}}                                      
(3.90, 0.0030, 32) & 1.1301(2) & 1.0615(3) & 1.234(3)  &  0.378(10)&  0.431(11) &0.410(11) &  0.214(9) \\	            
{\phantom{\huge{l}}}\raisebox{-.2cm}{\phantom{\Huge{j}}}                                      
(3.90, 0.0040, 32) & 1.1306(3) & 1.0621(3) & 1.238(6)  &  0.380(10)& 0.436(11) & 0.414(11) &  0.211(14)\\ \hline
{\phantom{\huge{l}}}\raisebox{-.2cm}{\phantom{\Huge{j}}}                            
(4.05, 0.0030, 32) & 0.9411(2) & 1.0518(6) & 1.215(7)  &  0.383(9)&  0.438(11) &0.415(10) &  0.224(6) \\			
{\phantom{\huge{l}}}\raisebox{-.2cm}{\phantom{\Huge{j}}}                            
(4.05, 0.0060, 32) & 0.9420(3) & 1.0534(5) & 1.240(10) &  0.383(9)& 0.436(11) & 0.412(11) & 0.231(7) \\			
{\phantom{\huge{l}}}\raisebox{-.2cm}{\phantom{\Huge{j}}}                            
(4.05, 0.0080, 32) & 0.9419(2) & 1.0519(4) & 1.218(9)  &  0.387(9)&  0.434(10) &0.408(10) & 0.226(6) \\ \hline	
{\phantom{\huge{l}}}\raisebox{-.2cm}{\phantom{\Huge{j}}}                            
(4.20, 0.0065, 32) & 0.7807(3) & 1.0479(4) & 1.222(8)  &  0.389(8)& 0.433(10) &0.421(9) & 0.234(12) \\			
{\phantom{\huge{l}}}\raisebox{-.2cm}{\phantom{\Huge{j}}}                            
(4.20, 0.0020, 48) & 0.7789(4)& 1.0463(6) & 1.209(5)  &  0.387(9)&  0.426(10) & 0.418(10) & 0.226(10) \\			
 \hline 
\end{tabular}}}
{\caption{\footnotesize  \label{tab:2} Detailed results for the hadronic quantities discussed in this paper, computed on each lattice data set specified in tab.~\ref{tab:1}.   }}
\end{table}

Hadron masses and decay constants are extracted from the study of the two-point correlation functions with operators chosen with desired quantum numbers, namely:  
\bea
J^{PC}=0^{-+} \qquad\qquad&& P= 2 \mu_c \ \bar c \gamma_5 c'  \,,\nn\\
J^{PC}=1^{--}  \qquad\qquad&& V_i= Z_A\  \bar c \gamma_i c'\quad{\rm or}\quad   T_{0i}= Z_T(\mu)\ \bar c \sigma_{0i} c'\,, \nn\\
J^{PC}=1^{+-}  \qquad\qquad&& T_{ij}= Z_T(\mu)\ \bar c \sigma_{ij} c'\qquad  i,j\in(1,2,3) \,,
\eea
In the above expressions the dependence of the renormalization constants on the bare lattice coupling is implicit, namely $Z_A\equiv Z_A(g_0^2)$, and  $Z_T(\mu)\equiv Z_T(g_0^2,\mu)$. Notice also that the above definition of the pseudoscalar operator $P$ is renormalization scale and scheme invariant both in the continuum and on the lattice with twisted mass QCD. To extract masses and decay constants one studies the large time separation between the operators in the two-point correlation functions. More specifically, 
\bea
\label{r1}
&& C_{P}(t)  =  \langle {\displaystyle \sum_{\vec x} }   P(\vec x; t) P^\dagger(0; 0) \rangle = - 4 \mu_c^2\sum_{\vec x} \langle {\rm Tr}\left[ S_c(\vec 0,0;\vec x,t)\gamma_5 S_c^\prime(\vec x,t;\vec 0,0) \gamma_5 \right] \rangle \nn\\
&&\qquad\qquad\qquad \xrightarrow[]{\displaystyle{ t\gg 0}} \; \frac{\cosh[  m_{\eta_c} (T/2-t)]}{ m_{\eta_c} }  \left| \langle 0\vert P(0)
\vert\eta_c (\vec 0) \rangle \right|^2 e^{- m_{\eta_c} T/2},\nn
\eea
\bea
&& C_{V}(t)  =  \langle {\displaystyle \sum_{\vec x} }   V_{i}(\vec x; t) V^\dagger_{i}(0; 0) \rangle = -Z_A^2\sum_{\vec x} \langle {\rm Tr}\left[ S_c(\vec 0,0;\vec x,t)\gamma_i S_c^\prime(\vec x,t;\vec 0,0) \gamma_i\right] \rangle \nn\\
&&\qquad \xrightarrow[]{\displaystyle{ t\gg 0}} \; \frac{\cosh[  m_{J/\psi} (T/2-t)]}{ m_{J/\psi} }  \left| \langle 0\vert V_i(0)
\vert J/\psi (\vec 0, \lambda) \rangle \right|^2 e^{- m_{J/\psi} T/2},\nn
\eea
\bea
&& C_{T}^{(-)}(t)  =  \langle {\displaystyle \sum_{\vec x} }   T_{0i}(\vec x; t) T_{0i}^\dagger(0; 0) \rangle = -Z_T^2\sum_{\vec x} \langle {\rm Tr}\left[ S_c(\vec 0,0;\vec x,t) \sigma_{0i}  S_c^\prime(\vec x,t;\vec 0,0)  \sigma_{0i} \right] \rangle  \nn\\
&&\qquad\qquad \xrightarrow[]{\displaystyle{ t\gg 0}} \; \frac{\cosh[  m_{J/\psi} (T/2-t)]}{ m_{J/\psi} }  \left| \langle 0\vert T_{0i}(0)
\vert J/\psi (\vec 0, \lambda) \rangle \right|^2 e^{- m_{J/\psi} T/2},\nn
\eea
\bea
&& C_{T}^{(+)}(t)  =  \langle {\displaystyle \sum_{\vec x} }   T_{ij}(\vec x; t) T_{ij}^\dagger(0; 0) \rangle = -Z_T^2\sum_{\vec x} \langle {\rm Tr}\left[ S_c(\vec 0,0;\vec x,t) \sigma_{ij}  S_c^\prime(\vec x,t;\vec 0,0)  \sigma_{ij} \right] \rangle \nn\\
&&\qquad\qquad \xrightarrow[]{\displaystyle{ t\gg 0}} \; \frac{\cosh[  m_{h_c} (T/2-t)]}{ m_{h_c} }  \left| \langle 0\vert T_{ij}(0)
\vert h_c (\vec 0, \lambda) \rangle \right|^2 e^{- m_{h_c} T/2}, 
\eea
where $i,j\in\{1,2,3\}$, and $T$ stands for the size of the periodic lattice in the time direction. 
Since our charmonia are taken to be at rest the matrix elements~(\ref{def1}) that appear in~(\ref{r1}) read:
\bea\label{r2}
&& \langle 0\vert P \vert\eta_c (\vec 0) \rangle =   f_{\eta_c} m_{\eta_c}^2 \,,\nn\\
&&\hfill\nn\\
&& \langle 0\vert V_i \vert J/\psi (\vec 0,\lambda)  \rangle = f_{ J/\psi} m_{J/\psi}   e_i^\lambda \,, \nn \\
&&\hfill\nn\\
&& \langle 0\vert T_{0i}(\mu) \vert J/\psi (\vec 0,\lambda)  \rangle = -i f_{ J/\psi}^T(\mu)  m_{J/\psi}   e_i^\lambda \,, \nn \\
&&\hfill\nn\\
&& \langle 0\vert T_{ij}(\mu) \vert h_c (\vec 0,\lambda)  \rangle = -i f_{ h_c}(\mu)  m_{h_c}  \varepsilon_{ijk}   e_k^\lambda \,.
\eea
In eq.~(\ref{r1}) we assumed the local source operators, which are needed for extraction of the decay constants. In practice, however, we implement the Gaussian smearing procedure in order to increase the overlap between the interpolating operator and the lowest state coupling to a given operator. The smearing procedure and the parameters used in actual computations have been discussed in refs.~\cite{ours,AA}.

The above matrix elements are then extracted by dividing the local-smeared and smeared-smeared correlation functions, where the coupling to the smeared correlation functions  can be studied from the smeared-smeared correlation functions in a way similar to eq.~(\ref{r1}). 
Hadron masses $a m_H$ ($H=\eta_c, J/\psi, h_c$) are extracted from the fit to a constant  on the plateau of the effective mass $m_{H}^{\rm eff}(t)$ defined from 
\bea
{\cosh\left[ m_{H}^{\rm eff}(t) \left( {\displaystyle{T\over 2}} - t\right)\right] \over \cosh\left[ m_{H}^{\rm eff}(t) \left( {\displaystyle{T\over 2}} - t -1\right)\right] }  = {C_{J}(t)\over C_{J}(t+1)}\,,
\eea  
with $J=P, V, T^{(+)}$ respectively. The results for the masses  have been presented in our previous paper~\cite{ours} and, for the reader's convenience, are presented in tab.~\ref{tab:2} of the present paper. The novelty is that we could also check that the results for the mass of $J/\psi$ state obtained from the correlation function $C_T^{(1)}(t)$ coincide with those we obtain from the study of $C_V(t)$ except that the errors are about $2\div 3$ times larger. Notice that only the mass of $m_{\eta_c}$ is given in the lattice units while the other masses are obtained from the fit to a constant of the ratios
\bea
R_{J/\psi}(t)= {m_{J/\psi}^{\rm eff}(t)\over m_{\eta_c}^{\rm eff}(t)}\,,\quad 
R_{h_c}(t)= {m_{h_c}^{\rm eff}(t)\over m_{\eta_c}^{\rm eff}(t)}\,,
\eea 
in which the statistical uncertainties cancel to a large extent. As for the decay constants, they are extracted in a way indicated in eq.~(\ref{r1}) and by using the definitions~(\ref{r2}). Their values are converted to physical units by using the lattice spacings quoted in tab.~\ref{tab:1}, and listed in  tab.~\ref{tab:2}. 

\begin{figure}
\vspace*{-0.8cm}
\begin{center}
\begin{tabular}{@{\hspace{-0.85cm}}cc}
{\resizebox{7.9cm}{!}{\includegraphics{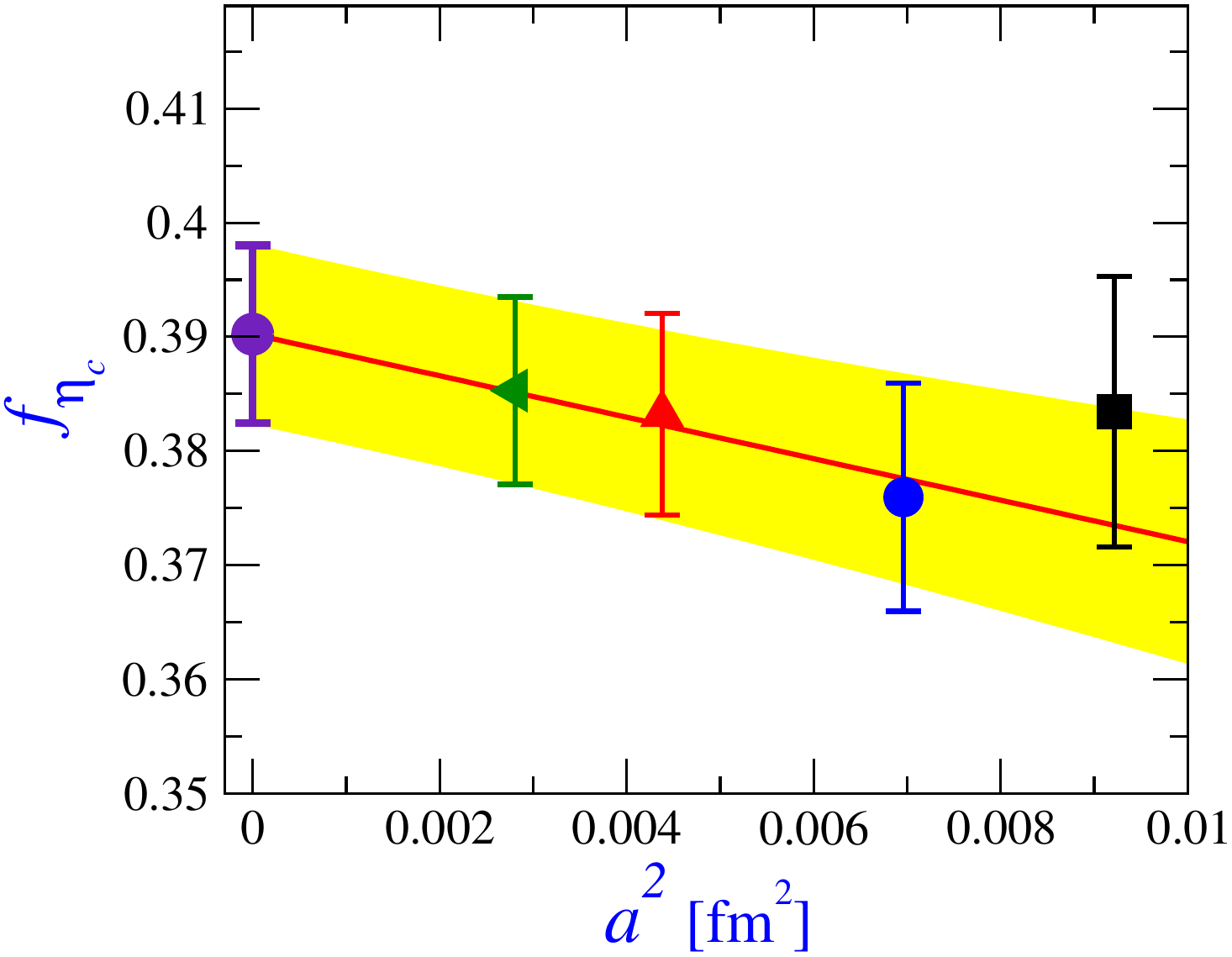}}}&{\resizebox{7.9cm}{!}{\includegraphics{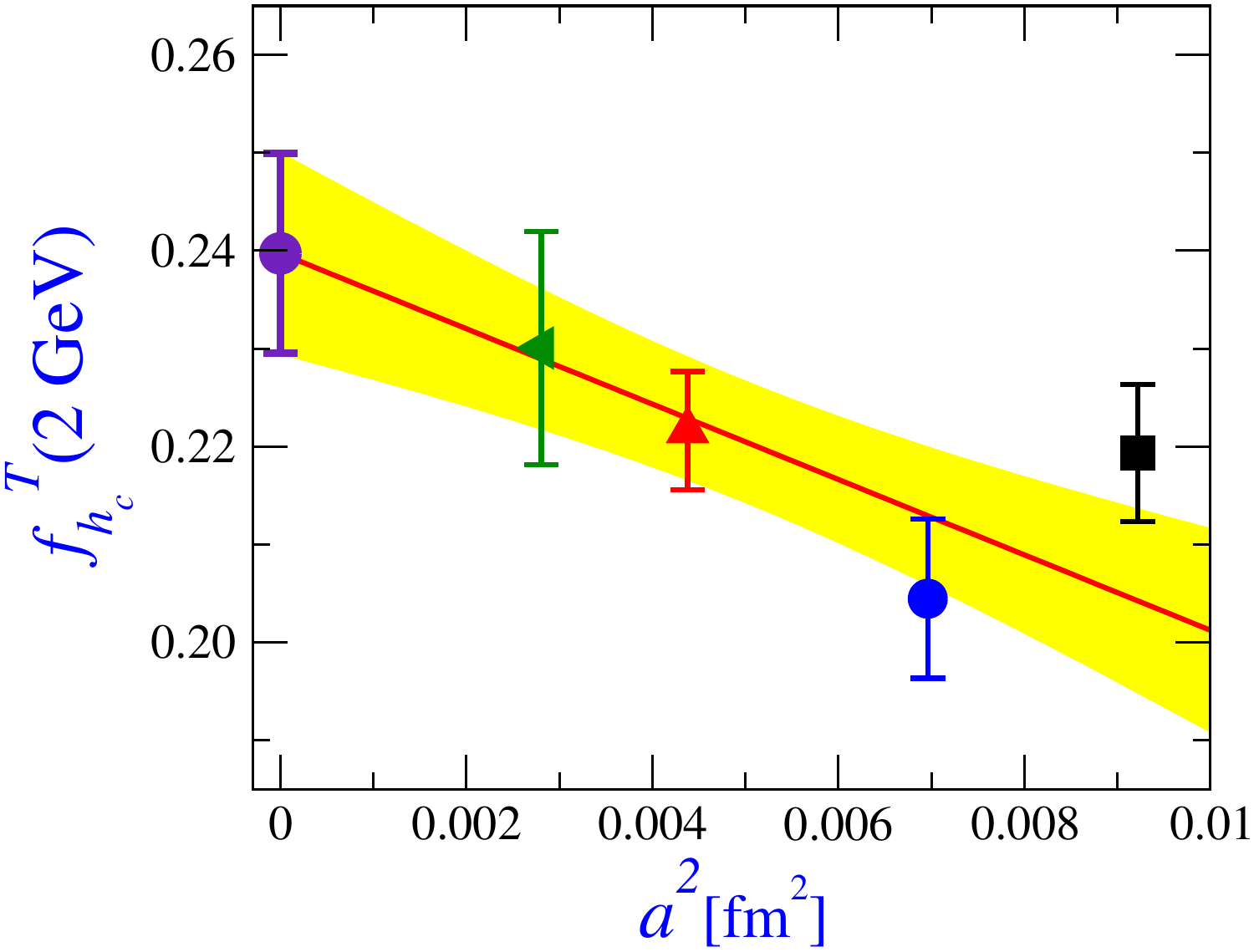}}} \\
{\resizebox{7.7cm}{!}{\includegraphics{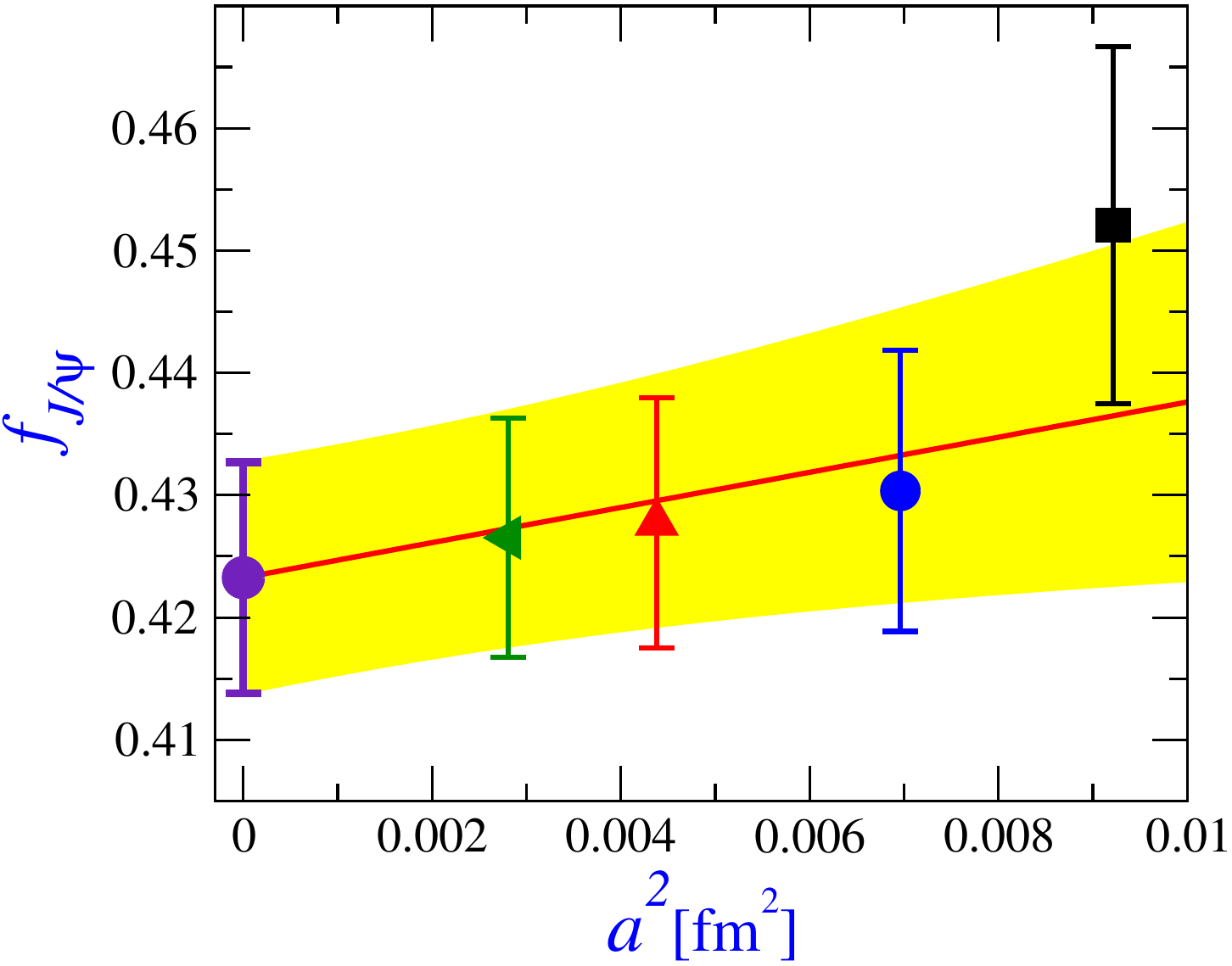}}}&{\resizebox{7.7cm}{!}{\includegraphics{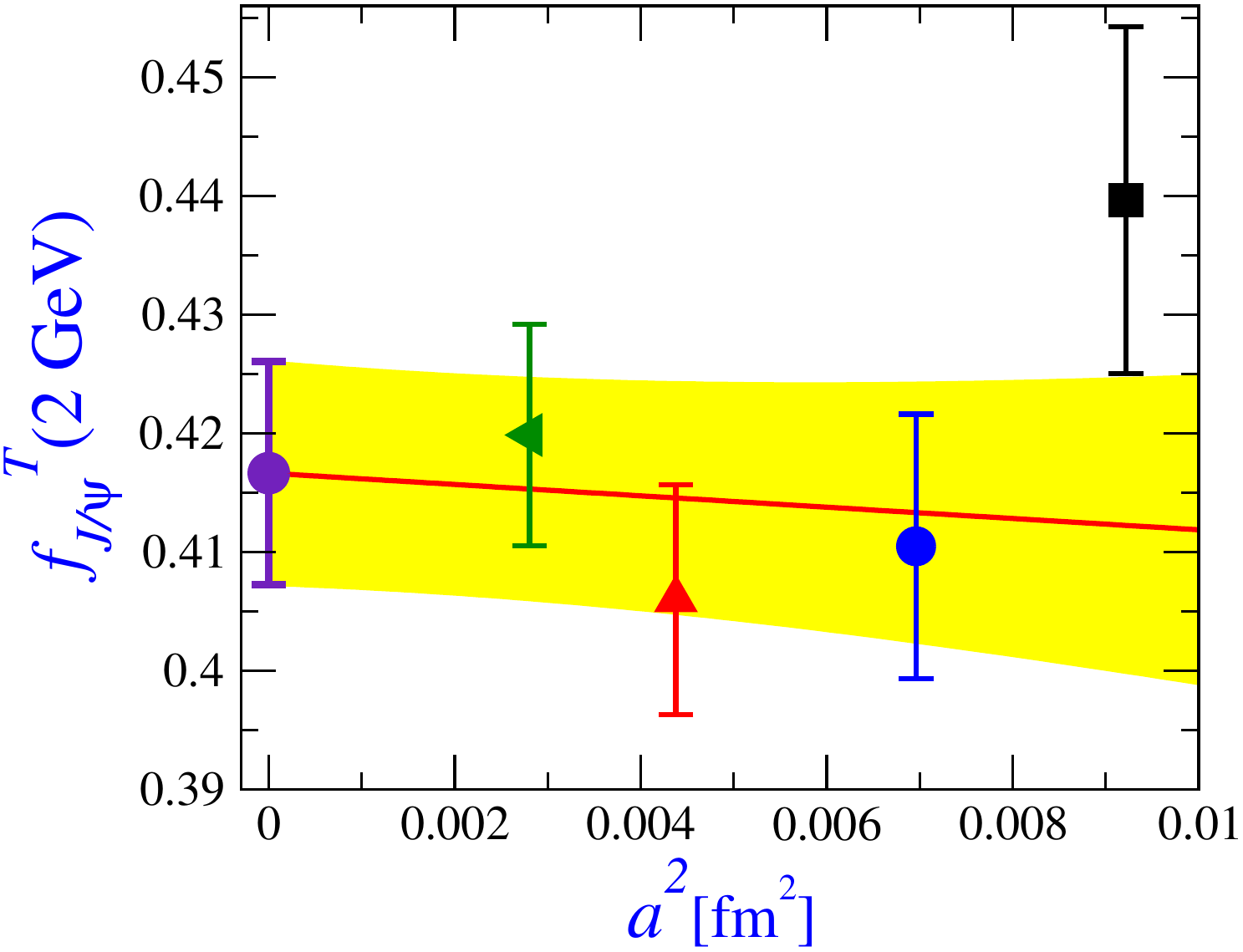}}} \\
\end{tabular}
\vspace*{-.1cm}
\caption{\label{fig:2}{\footnotesize 
Continuum extrapolation of $f_{\eta_c}$, $f_{h_c}(2\ \gev)$, $f_{J/\psi}$, and $f_{J/\psi}^T(2\ \gev)$. Yellow bands correspond to the continuum extrapolation made by using eq.~(\ref{eq:cont}) without including the results obtained from the coarse lattices ($\beta=3.8$).
 } }
\end{center}
\end{figure}

\begin{figure}[h!]
\begin{center}
\begin{tabular}{@{\hspace{-0.85cm}}cc}
{\resizebox{8cm}{!}{\includegraphics{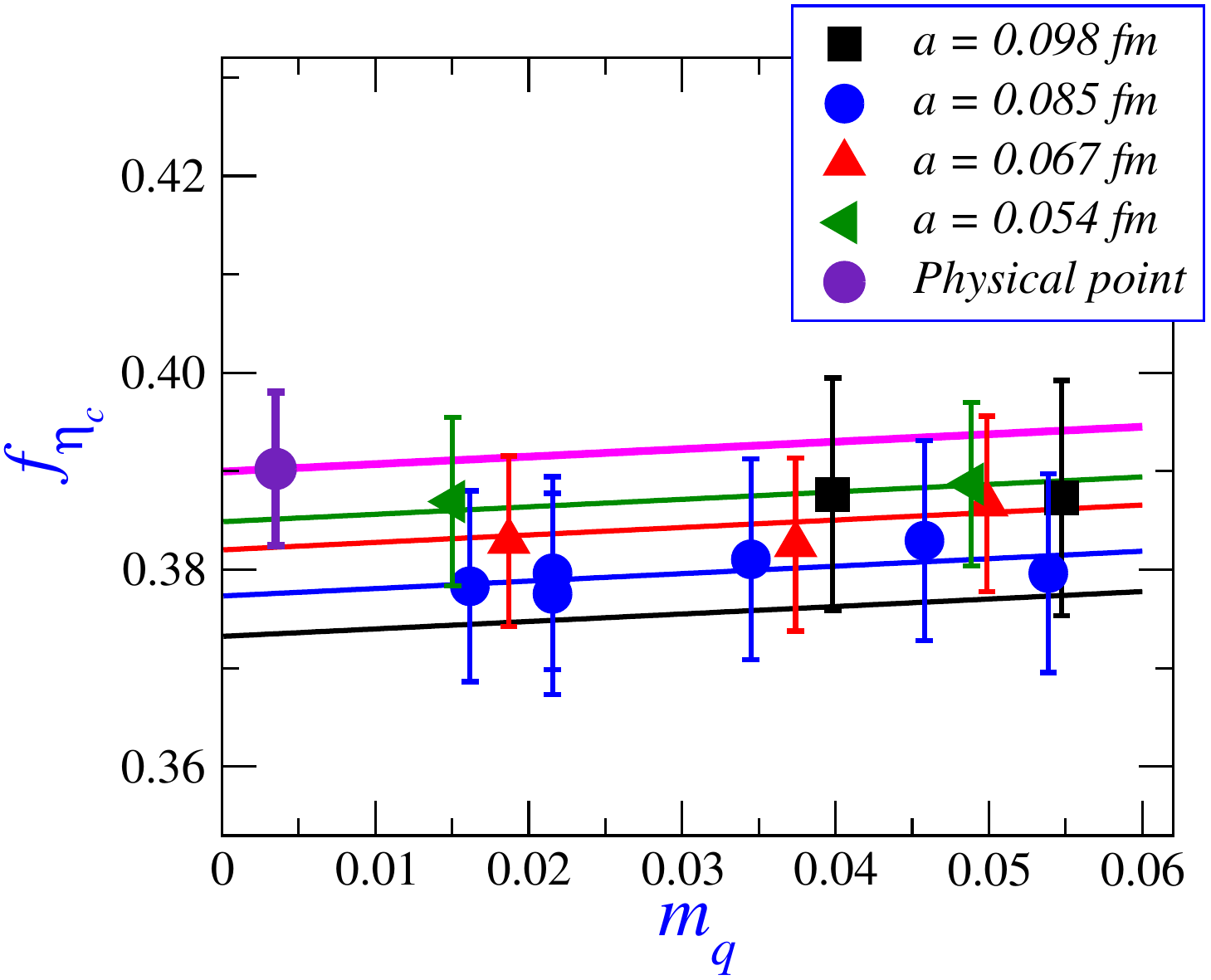}}}&{\resizebox{8cm}{!}{\includegraphics{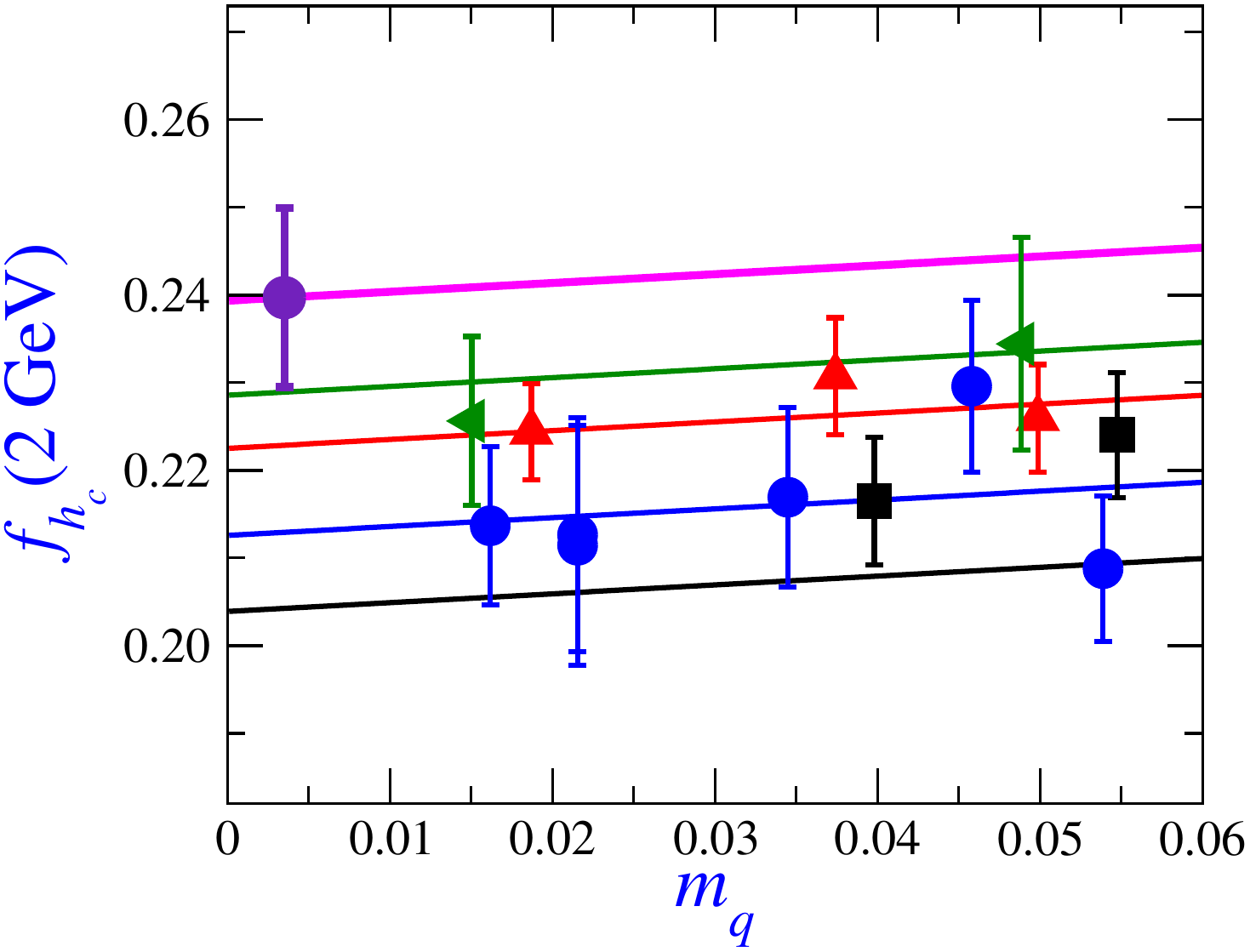}}} \\
{\resizebox{8cm}{!}{\includegraphics{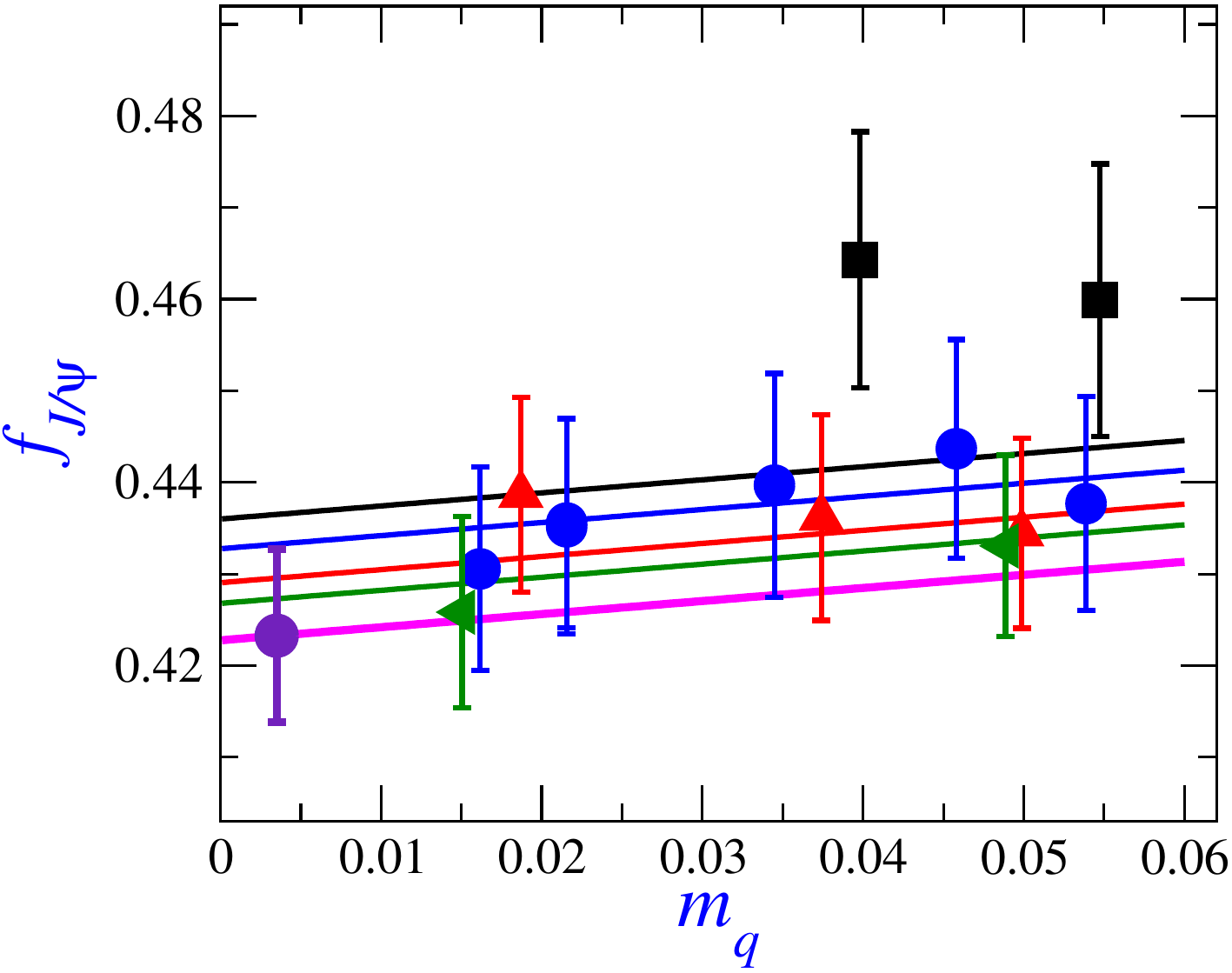}}}&{\resizebox{8cm}{!}{\includegraphics{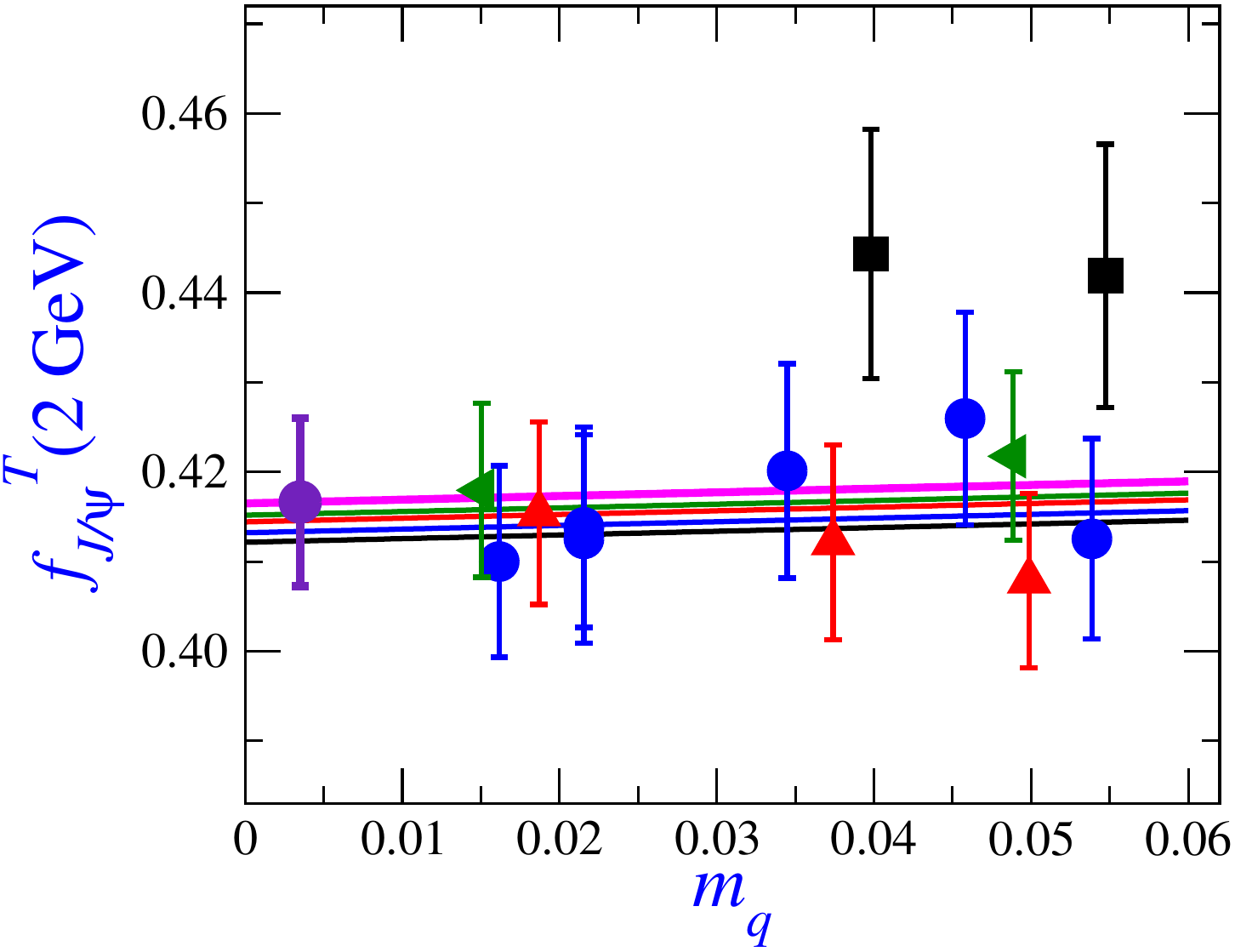}}} \\
\end{tabular}
\vspace*{-.1cm}
\caption{\label{fig:1}{\footnotesize 
Dependence of  $f_{\eta_c}$, $f_{h_c}(2\ \gev)$, $f_{J/\psi}$, and $f_{J/\psi}^T(2\ \gev)$ on the sea quark mass $m_q\equiv m_q^\msbar(2\ \gev)$ at each of our lattice spacings, as well as in the continuum limit. Separation among the curves, obtained from the simultaneous fit of our data to eq.~(\ref{eq:cont}),  indicates the dependence on the finite lattice spacing already shown in fig.~\ref{fig:2}.  All quantities are displayed in physical units (in GeV). 
 } }
\end{center}
\end{figure}

To reach a physically interesting results we need to extrapolate our decay constants, obtained at four lattice spacings, to the continuum limit. Since the physical quantities computed with maximally twisted mass QCD on the lattice are automatically ${\cal O}(a)$ improved, the leading terms are those proportional to $a^2$. Furthermore, at each lattice spacing we computed the charmonium couplings for several values of the dynamical light quark masses, which is useful in order to check on their dependence on the {\sl sea} quark mass, like we did in our previous paper where we showed that the masses of charmonia are completely insensitive to the sea quark mass~\cite{ours}. To get the physically relevant result in the continuum limit, the decay constants from tab.~\ref{tab:2} are therefore fit to the following form, 
\bea\label{eq:cont}
f_H = f_H^{\rm cont.} \left[ 1 +  b_{H} m_{q} + c_{H} {a^2\over (0.086\ {\rm fm})^2}\right]\,,
\eea
where the parameter $b_{H}$ measures the dependence on the sea quark mass, denoted by $m_q\equiv m_q^\msbar(2\ \gev)$, while the parameter $c_{H}$ measures the leading discretization effects. Division by $a_{\beta=3.9}=0.086$~fm is made for convenience. The linear fit~(\ref{eq:cont}) in $a^2$ provides an adequate description of all our results if we leave out the data obtained at $\beta=3.80$, as it can be appreciated from the plots provided in fig.~\ref{fig:2}. Note however that the extrapolation curve shown in fig.~\ref{fig:2} takes into account the fact that at each lattice spacing the results are obtained for several values of the sea quark mass. The dependence on the sea quark mass is shown in fig.~\ref{fig:1}. We therefore report the results of the fit to eq.~(\ref{eq:cont}) separately for the case in which the data at $\beta=3.8$ are left out, and the results of the continuum extrapolation with all the lattice data included, cf. tab.~\ref{tab:3}. Although the quality of the fit deteriorates when all the lattice data are included, its $\chi^2$/d.o.f. is still acceptable and we prefer to use both results and include the difference in the estimate of the systematic uncertainty. That leads us to our final estimates: 
\begin{align}\label{eq:latt-results}
& f_{\eta_c}=387(7)(2)~\mev\,,&& f_{J/\psi}=418(8)(5)~\mev\,,\nn\\
&  f_{J/\psi}^T(2\ \gev)=410(8)(6)~\mev\,,&& f_{h_c}(2\ \gev)=235(8)(5)~\mev\,.
\end{align}

\begin{table}[t!!]
{\hspace{-0.5cm}}{\scalebox{.91}{\begin{tabular}{|c|ccc| ccc|}  \cline{2-7}
\multicolumn{1}{l|}{} & \multicolumn{3}{c|}{without $\beta=3.8$} & \multicolumn{3}{c|}{all lattices}   \\   \hline
{\phantom{\huge{l}}}\raisebox{-.2cm}{\phantom{\Huge{j}}}
$f_H$ &  value   &  $b_H$ & $c_H$ &  value   &  $b_H$ & $c_H$   \\   \hline  \hline
{\phantom{\huge{l}}}\raisebox{-.2cm}{\phantom{\Huge{j}}}
$f_{\eta_c}$&  $390(8)~\mev$ & $0.19(13)~\gev^{-1}$ & $-0.03(1)$ &  $385(7)~\mev$ & $0.32(12)~\gev^{-1}$ & $-0.02(1)$ \\ 
{\phantom{\huge{l}}}\raisebox{-.2cm}{\phantom{\Huge{j}}}
$f_{J/\psi}$&  $423(9)~\mev$ & $0.3(2)~\gev^{-1}$ & $+0.02(2)$ &  $414(8)~\mev$ & $0.5(2)~\gev^{-1}$ & $0.05(2)$  \\ 
{\phantom{\huge{l}}}\raisebox{-.2cm}{\phantom{\Huge{j}}}
$f_{J/\psi}^T(2\ \gev)$&  $416(9)~\mev$ & $0.1(2)~\gev^{-1}$ & $-0.01(2)$ &  $403(8)~\mev$ & $0.4(2)~\gev^{-1}$ & $0.03(2)$  \\ 
{\phantom{\huge{l}}}\raisebox{-.2cm}{\phantom{\Huge{j}}}
$f_{h_c}(2\ \gev)$&  $239(10)~\mev$ & $0.4(6)~\gev^{-1}$ & $-0.11(5)$ &  $230(6)~\mev$ & $0.7(6)~\gev^{-1}$ & $-0.07(2)$  \\ 
 \hline
\end{tabular}}}
{\caption{\footnotesize  \label{tab:3} Results of the fit of our data presented in tab.~\ref{tab:2} to a form given in eq.~(\ref{eq:cont}) without/with the results at $\beta =3.8$ included in the fit.}}
\end{table}

Two of the couplings discussed in this paper ($f_{J/\psi}$ and $f_{\eta_c}$) have been already computed on the lattice in an unquenched setup but with the different lattice regularization. By using the staggered quark action and by including $\nf=2+1$ dynamical light flavors, the authors of ref.~\cite{Davies:2010ip} obtained $f_{\eta_c}=395(2)~\mev$, in the continuum limit. 
With a similar setup, the same collaboration later reported $f_{J/\psi}=405(6)(2)~\mev$~\cite{Donald:2012ga}. Knowing that the lattice actions are very different, the fact that these results agree quite well in the continuum limit with our values~(\ref{eq:latt-results}), is a good indication of the robustness of the lattice QCD predictions. Our results indicate that there is no dependence of the charmonium quantities (masses decay constants and the form factors discussed in ref.~\cite{ours}) on the light sea quark mass. The results presented in refs.~\cite{Davies:2010ip,Donald:2012ga} also suggest that the decay constants $f_{\eta_c,J/\psi}$ do not depend on the strange sea quark mass. Finally, we remark that the values for $f_{J/\psi}^T$ and $f_{h_c}$ are new.

\section{\label{sec-3} Phenomenology}

In this section we comment on two topics of phenomenological interest, already mentioned in introduction of the present paper, namely the $\eta_c\to \gamma\gamma^{(\ast)}$ decay, and the factorization of the non-leptonic $B$-decays to two mesons, one of which is a charmonium.

\subsection{$\eta_c\to \gamma\gamma^{(\ast)}$}


For a theoretical estimate of $\Gamma(\eta_c\to \gamma\gamma)$ the non-perturbative information is essential and is related to $f_{\eta_c}$. 
In the standard derivation of the expression for $\Gamma(\eta_c\to \gamma\gamma)$ one starts from the $\eta_c\to \gamma^\ast \gamma^\ast$ decay amplitude,
\bea
{\cal A}= i\ 4\pi \alpha_{\rm em}  \left(\frac{2}{3}\right)^2  F(k_1^2,k_2^2) \ \varepsilon_{\mu\nu\alpha\beta}\  e_1^\alpha e_2^\beta k_1^\mu k_2^\nu\,,
\eea
where $k_{1,2}$ and $e_{1,2}$ stand for the momenta and polarization vectors of the two off-shell photons. One then assumes the validity of factorization of the soft QCD dynamics of $\eta_c$ and the hard rescattering of $c\bar c$ into photons. By taking one photon on-shell the other is expected to be energetic enough for factorization to be applicable. The resulting process $\eta_c\to \gamma\gamma^\ast$ is then described by the form factor $F_{\gamma\eta_c}(q^2) \equiv F(q^2,0)$ which enters directly the expression for the $\eta_c\to \gamma\gamma$ decay rate as,
\bea\label{eq:gammagamma}
\Gamma(\eta_c\to \gamma\gamma) = {4\pi \alpha_{\rm em}^2\over 81} m_{\eta_c}^3 \vert F_{\gamma\eta_c}(0)\vert^2\,.
\eea
The form factor $F_{\gamma\eta_c}(Q^2)$ can be studied experimentally through $d\sigma(e^+e^-\to e^+e^-\eta_c)/dQ^2$ ($Q^2=-q^2>0$), a process driven by $\gamma\gamma^\ast \to \eta_c$.  
In this way, after a detailed measurement of such a process, the BaBar Collaboration was able to determine the shape of $F_{\gamma\eta_c}(Q^2)$ in a large energy window corresponding to  $Q^2\in (0, 50)\ \gev^2$~\cite{Lees:etac}. They found that the data are very well described by a single pole form, with the pole being at $m_{\rm pole}=2.9(1)(1)$~GeV. Such a pole-like behavior was predicted by (quenched) QCD on the lattice~\cite{Dudek:2006ut}, and is compatible with the vector meson dominance. As for the intercept of the form factor, $ F_{\gamma\eta_c}(0)$, different models give different answers~\cite{models}. For example, the perturbative QCD approach of ref.~\cite{kroll}, results in 
\bea\label{eq:feldmann}
F_{\eta_c\gamma}(0) \simeq { 4 f_{\eta_c}\over m_{\eta_c}^2 + 2 \langle {\mathbf{k}}_\perp^2\rangle }\,,
\eea
where $\sqrt{ \langle {\mathbf{k}}_\perp^2\rangle}$ is the mean transverse momentum of the $c$-quark with respect to the momentum of $\eta_c$. After assuming  $\langle {\mathbf{k}}_\perp^2\rangle =0$, one gets the expression usually employed in the literature~\cite{models}. Similarly, the authors of ref.~\cite{lansberg} used the heavy quark spin symmetry to estimate $\Gamma(\eta_c\to \gamma\gamma)$, and their expression for the form factor coincides with  eq.~(\ref{eq:feldmann}) after replacing $2 \langle {\mathbf{k}}_\perp^2\rangle \to   b_{\eta_c}m_{\eta_c}$, with $b_{\eta_c}= 2 m_c-m_{\eta_c}$. The latter quantity  is clearly ambiguous as the quark mass is a renormalization scale and scheme dependent quantity. If one assumes $m_c$ to be the pole mass, the value of   $b_{\eta_c}$ can be fixed if one knows $ f_{\eta_c}$ and $\Gamma^{\rm exp.}(\eta_c\to \gamma\gamma)$. Taking $b_{\eta_c}=0$  ($\to  \langle {\mathbf{k}}_\perp^2\rangle = 0$) reduces eq.~(\ref{eq:feldmann}) to the formula most frequently used in the literature. 

In ref.~\cite{melikhov}, by imposing the local quark-hadron duality on the decay amplitudes, the authors derived a yet another expression for $F_{\eta_c\gamma}(0)$, namely
\bea
F_{\eta_c\gamma}^{\rm LD}(0)=  { 6 f_{\eta_c}\over m_{V}^2 }\,,
\eea
where $m_V=3.75(25)$~GeV has been fixed from the single pole fit to the BaBar data at large values of $Q^2$. In all these expressions $f_{\eta_c}$ enters decisively and its impact on eq.~(\ref{eq:gammagamma}) should be checked against the experimental data. Another possibility is to rely on the nearest vector meson dominance (VMD) hypothesis, namely~\cite{swanson1},
\bea\label{eq:VMD}
F_{\eta_c\gamma}^{\rm VMD}(0)=  2 {  f_{J/\psi}\over m_{J/\psi} } \  {  2 V^{J/\psi\to \eta_c}(0)\over m_{J/\psi} + m_{\eta_c} } \,,
\eea
which can be tested since the value of the $J/\psi\to \eta_c\gamma$  form factor is nowadays known from the lattice QCD studies of refs.~\cite{ours,Donald:2012ga}.

By using eq.~(\ref{eq:gammagamma}) or (\ref{eq:feldmann})  we can write
\bea\label{eq:rategg}
\Gamma(\eta_c\to \gamma\gamma) = {64\pi \alpha_{\rm em}^2\over 81 m_{\eta_c}}   {f_{\eta_c}^2 \over (1+\delta )^2}\,.
\eea
With the experimentally established ${\cal B}(\eta_c\to \gamma\gamma)=(1.57\pm  0.12)\times 10^{-4}$, and $\Gamma(\eta_c)=32.0(9)$~MeV, we have $\Gamma^{\rm exp}(\eta_c\to \gamma\gamma)= 5.0(4)$~keV, which together with our $f_{\eta_c}= 0.387(8)$~GeV, allows us to deduce the value of  $\delta=0.15(5)\ \gev^2$. That value is too large to be interpreted as $\sqrt{\langle {\mathbf{k}}_\perp^2\rangle}=0.81(14)$~GeV,~\footnote{Even if one assumes this value to be correct, then one could not fit the BaBar data at large $Q^2$'s by using the expression $F_{\eta_c\gamma}(Q^2)/F_{\eta_c\gamma}(0)=1/(Q^2+m_{\eta_c}^2 + 2 \langle {\mathbf{k}}_\perp^2\rangle)$~\cite{kroll}, where the perturbative approach is expected to work better.} and also too large to be identified as $b_{\eta_c}=\delta m_{\eta_c}=0.46(16)$~GeV. 
Finally, we should say that the VMD is actually quite good an approximation. By using $V^{J/\psi\to \eta_c}(0)= 1.92(3)(2)$ computed in ref.~\cite{ours}, together with our result for $f_{J/\psi}$, inserted in eq.~(\ref{eq:VMD}), for the di-photon decay width we get $\Gamma(\eta_c\to \gamma\gamma) =6.0(4)$~keV.~\footnote{This number remains essentially unchanged if we used the lattice QCD results obtained in ref.~\cite{Donald:2012ga}, $V^{J/\psi\to \eta_c}(0)= 1.90(7)(1)$ and $f_{J/\psi}=0.405(6)$~MeV. We get $\Gamma(\eta_c\to \gamma\gamma) =5.9(5)$~keV.}

We conclude that the usual expression for $\Gamma(\eta_c\to \gamma\gamma)$ based on factorization approximation [$\delta=0$ in eq.~(\ref{eq:rategg})] leads to the result larger than the experimental value:   $\Gamma^{\rm fact.}(\eta_c\to \gamma\gamma)=(6.64\pm 0.27)$~keV, vs. $\Gamma^{\rm exp.}(\eta_c\to \gamma\gamma)=(5.0\pm 0.4)$~keV. That discrepancy can be studied in a systematic way by means of non-relativistic QCD expansion, along the lines of ref.~\cite{nora-antonio}. Research in this direction, to elucidate the origin of this discrepancy, would be welcome. 

\subsection{Non-leptonic $B$ decays to charmonia}

By using the factorization approximation, the decay rate of the Class-II non-leptonic $B$-decays in the Standard Model can be written as
\begin{align}\label{eq:rates}
\Gamma (B\to J/\psi K) &= {G_F^2 \vert V_{cb}V_{cs}^\ast\vert^2 \over 32 \pi m_B^3} \ \lambda^{3/2}(m_B^2,m_{J/\psi}^2,m_K^2)\ a_2^2\  f_{J/\psi}^2 \left[ f_+^{B\to K}(m_{J/\psi}^2)\right]^2\,,\nn \\
\Gamma (B\to \eta_c K) &= {G_F^2 \vert V_{cb}V_{cs}^\ast\vert^2 \over 32 \pi m_B^3} \ (m_B^2-m_K^2)^2 \lambda^{1/2}(m_B^2,m_{\eta_c}^2,m_K^2)\ a_2^2\  f_{\eta_c}^2 \left[ f_0^{B\to K}(m_{\eta_c}^2)\right]^2\,,
\end{align}
where the coefficient $a_2$ is a combination of Wilson coefficients computed in perturbation theory, encoding the information about the short distance physics. That quantity is considered as a parameter in the generalized factorization~\cite{neubert-stech}, that is to be obtained from the experimentally measured one decay mode and then used to describe the other modes of the given Class. By taking the ratios of the above rates, we get 
\bea\label{eq:factII}
{B(B\to \eta_c K)\over B(B\to J/\psi K)} = {(m_B^2-m_K^2)^2 \lambda^{1/2}(m_B^2,m_{\eta_c}^2,m_K^2)\over \lambda^{3/2}(m_B^2,m_{J/\psi}^2,m_K^2)} \ \left( {f_{\eta_c}\over f_{J/\psi}}
\right)^2 \ \left( {f_0^{B\to K}(m_{\eta_c}^2) \over f_+^{B\to K}(m_{J/\psi}^2)}
\right)^2\,,
\eea
where $\lambda(a,b,c)= [a^2 - (b+c)^2] [a^2 - (b-c)^2]$. 
With our result $f_{\eta_c}/f_{J/\psi}=0.926(6)$, one can then compare the measured charged and neutral $B$-decay modes~(\ref{eq:00}) with eq.~(\ref{eq:factII})  and deduce,
\bea
{f_+^{B\to K}(m_{J/\psi}^2) \over f_0^{B\to K}(m_{\eta_c}^2)} = \left. 1.53(10)\right|_{B^\pm-\rm mode},\left. 1.56(13)\right|_{B^0-\rm mode}.
\eea
These results are consistent with $\approx 1.44$, as obtained from the QCDSR calculation near the light cone in ref.~\cite{BallZwicky,DuplancicMelic}. They are also consistent with $1.51(3)$ obtained in the quenched lattice QCD study of ref.~\cite{withnejc}, but not as well with $1.37(2)$ recently obtained in the unquenched lattice study with non-relativistic QCD employed to treat the heavy quark~\cite{bouchard}. So this information can be used either to get an idea on the above ratio of the form factors, or as a measure of the deviation with respect to the factorization approximation if the form factors are taken from elsewhere.

\section{\label{sec-4}Summary}

In this paper we presented results of our analysis of four decay constants of the charmonium states. By adopting the strategy of ``one resonance + continuum" in the moment QCDSR analysis, we found that the values of the decay constants $f_{J/\psi}$ and $f^T_{J/\psi}$ agree quite well with those obtained through the simulations of QCD on the lattice, in the continuum limit. On the other side the QCDSR results for the pseudoscalar meson decay constant $f_{\eta_c}$ are lower than those obtained on the lattice. Similar holds true for $f_{h_c}$, decay constant of the recently observed $J^{PC}=1^{+-}$ charmonium state.  
Adding more states to the hadronic side of the sum rules helps improving the stability of the sum rules, while the value of the decay constant remains practically unchanged. One reason for disagreement of the QCDSR estimate of  $f_{\eta_c}$ with that obtained on the lattice might be related to the fact that the non-perturbative contribution to the sum rules, proportional to the gluon condensate, has been fixed from the detailed analysis of the vector-vector correlation function. A possible explanation of that discrepancy is that the series of power corrections is truncated and that the higher order terms affect different correlation function differently, which is why $f_{\eta_c}$ and $f_{h_c}$ are not as well reproduced by the QCSR as it is the case with $f_{J/\psi}$ and $f^T_{J/\psi}$. We plan to come back to that issue in the near future.  
In the one resonance plus continuum setup, the expected precision of the moment QCDSR estimates is of the order $10\div 15\%$,  which is what we observe with our results. Note also that the results presented in this paper for the spectral function and for the gluon condensate contributions to the correlation function of tensor densities are new. We should stress that in view of the approximations made in the method of QCDSR, the agreement of $f_{J/\psi}$, $f_{J/\psi}^T$ and even $f_{\eta_c}$ with the results obtained from lattice QCD is quite remarkable. The case of $f_{h_c}$ is an exception, however. We did not attempt to remedy that discrepancy by adding an extra term to the series of power corrections but we plan to come back to that issue in the future.

Our lattice computation of the same set of decay constants is made in the Wilson regularization of QCD by including the maximally twisted mass term, with $\nf=2$ dynamical light quarks included in the gauge field configurations. From the simulations made at four different lattice spacings we were able to take the continuum limit. We find that the charmonium decay constants are insensitive to the variation of the mass of the light dynamical quarks. Non-perturbatively computed renormalization constants were implemented in our computation, and our final results are: 
\begin{align}
& f_{\eta_c}=(387\pm 7)~\mev\,,&& f_{J/\psi}=(418\pm 9)~\mev\,,\nn\\
&  f_{J/\psi}^T(2\ \gev)=(410\pm 10)~\mev\,,&& f_{h_c}(2\ \gev)=(235\pm 9)~\mev\,,\nn
\end{align}
where we combined the statistical and systematic errors in the quadrature. 

With the above results in hands we were able to address two issues of phenomenological interest. First, and by using our $f_{\eta_c}$, we get that the standard formula for the decay width of $\eta_c\to \gamma\gamma$,  does not reproduce the experimentally measured width, which might be an indication of the presence of non-factorizable terms. With our values for  $f_{\eta_c}/ f_{J/\psi}$ we were able to check on the factorization approximation in the Class-II non-leptonic decays of $B$-mesons. We found that the most recent lattice results for the $B\to K$ indicate the violation of the factorization approximation, whereas those obtained by the QCDSR near the light cone as well as the older lattice results are quite consistent with what we extracted for  ${f_+^{B\to K}(m_{J/\psi}^2)/f_0^{B\to K}(m_{\eta_c}^2)}$ from the ratios of non-leptonic decay channels together with our $f_{\eta_c}/ f_{J/\psi}$. Another lattice QCD estimate of this ratio of form factors would be highly welcome.

Finally, we would like to emphasize that the results for $R^T_{J/\psi}=f^T_{J/\psi}(2\ \gev)/f_{J/\psi}$ as obtained in our QCDSR analysis agree very well with those computed on the lattice:
\bea
R^T _{J/\psi} = \biggl.0.975\pm 0.010\biggr|_{\rm QCDSR},  \biggl.\quad 0.981\pm 0.008\biggr|_{\rm lattice QCD}.
\nn
\eea

\section*{Acknowledgments}
We thank the members of the ETM Collaboration for making their gauge field configurations publicly available. Numerical computations are performed using the HPC resources of IDRIS Orsay, thanks to the computing time given to us by GENCI (2013-056808). We also thank C.~Bouchard, A.~Khodjamirian and A.~Radyushkin for correspondence related to their respective works.
\newpage

\section*{Appendix: Spectral Functions and gluon condensate contributions}

\begin{figure}[htb]
\begin{center}
{\resizebox{2.9cm}{!}{\includegraphics{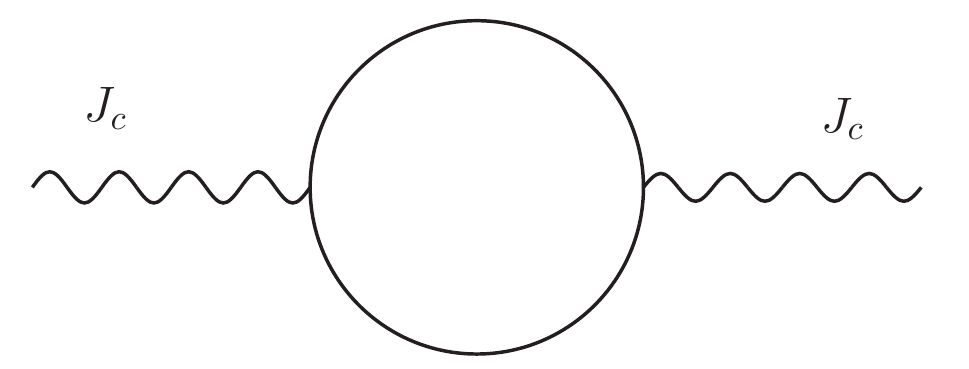}}}\hspace*{1cm}{\resizebox{8.9cm}{!}{\includegraphics{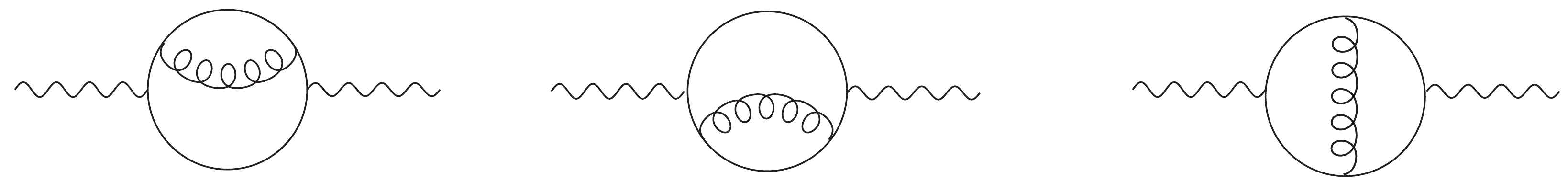}}} 
\caption{\label{calc1}{\footnotesize 
LO and NLO contributions to the spectral functions.
 } }
\end{center}
\end{figure}

While the leading contributions to the spectral functions, $\rho_i^{\rm pert}(s) $, are easy to calculate, the ${\cal O}(\alpha_s)$ corrections are quite demanding as they require evaluating the two-loop diagrams. To derive  perturbative spectral functions $\rho^{\rm pert}_i(s)={\rm Im }\Pi_i^{\rm pert}/\pi$ one needs to calculate  the imaginary part of the diagrams shown in fig.~\ref{calc1}, 
with both external currents being either $V_\mu =\bar c\gamma_\mu c$, or $P = 2 m_c\  i\bar c\gamma_5 c$, or $T_{\mu\nu}=\bar c\sigma_{\mu\nu}c$. 
By using the standard approach, i.e. multiplying by appropriate projectors and expressing the scalar products 
in numerators in terms of those in denominators, one performs the tensor decomposition to the basic scalar Feynman integrals.  
The calculation of the relevant two-loop scalar integrals could be challenging, but since we are only interested in their imaginary part 
 the task becomes much simpler. We computed the scalar integrals in two ways: (i) by the `cut'-technique using 
the Cutkosky rules and (ii) by a directly extracting the  imaginary part of the integrals from their Feynman parameter representation. 
Both ways lead to the same results. Since we used dimensional regularization, the above mentioned calculations were
performed in $d$-dimensions. Finally, besides the renormalization of $\alpha_s$, $m_c$ and 
the quark field, we accounted for the renormalization of the operators in the $\msbar$ scheme. 
The above choice of the pseudoscalar density $P$ is particularly convenient because the anomalous dimension of the $i\bar c \gamma_5 c$ cancels against that of the quark mass, so that  $P$ is renormalization group invariant.  Therefore, the only correlator in which one should take care of the anomalous dimension is that involving the tensor densities. A standard procedure consists in connection the bare and renormalized 
current via 
\bea
j_B(x) = Z_j^{-1} Z_2\  j_R\,,
\eea
where $Z_2$ is the quark field renormalization constant and the anomalous dimension is derived as 
\bea
\gamma_j = \mu^2 \frac{d}{d\mu^2} \ln \left ( \frac{Z_j}{Z_2} \right )\,,
\eea 
In the expansion 
\bea
\gamma_j =  \gamma_j^{(0)} \frac{\alpha_s}{\pi} + \; ....
\eea
it follows that $\gamma_T^{(0)} = 2/3$. Therefore after renormalization, the constants $f_{h_c}$ and $f_{J/\psi}^T$ are $\mu$-dependent quantities.   


In the following we give the full expressions for perturbative spectral functions, $\rho^{\rm pert}_i(s)\equiv {\rm Im}\Pi_i^{\rm pert}(s)/\pi$ ($i=P,V,+,-$), 
written separately for the leading and the next-to-leading term in $\alpha_s$, namely,  
\bea
\rho_i^{\rm pert}(s) =   \rho_i^{(0)}(s)+{\alpha_s\over\pi}\rho_i^{(1)}(s)\,.
\eea
and
\begin{align}
\rho_P^{(0)}(s) &= {3\over 8\pi^2} s v\,, \nn\\
\rho_P^{(1)}(s) &= \frac{3}{8\pi^2} s \frac{3 v^2-1}{v} \ln \frac{\mu^2}{\bar{m}^2} \nonumber \\
& + \frac{s}{2\pi^2} \left[B(v) + \left(\frac{19}{16} + \frac{2}{16}v^2 + \frac{3}{16} v^4 \right)\ln\frac{1+v}{1-v} - 
\frac{1}{v} + \frac{45}{8} v - \frac{3}{8}v^3\right],\\
%
\rho_V^{(0)}(s)& = {1\over 8\pi^2}  v \left(3-v^2\right)\,,\nn\\
\rho_V^{(1)}(s) &= - \frac{3}{8\pi^2} \frac{(1-v^2)^2}{v} \ln \frac{\mu^2}{\bar{m}^2} \nonumber \\
& \quad + \frac{1}{2\pi^2} \left[\left(1-\frac{1}{3}v^2\right)B(v) + \left(\frac{33}{24} + \frac{22}{24}v^2 - 
\frac{7}{24} v^4 \right)\ln\frac{1+v}{1-v} - \frac{1}{v} + \frac{39}{12} v - \frac{21}{12}v^3\right],\\
\rho_+^{(0)}(s) &=  {1\over 8\pi^2} v^3\nn\\
\rho_+^{(1)}(s) &=  \frac{3}{8\pi^2} v \left(v^2 -1\right) \ln \frac{\mu^2}{\bar{m}^2} \nonumber \\
&+ \frac{1}{6\pi^2}  \left[v^2 B(v) + \left(\frac{13}{16} + \frac{28}{16}v^2 + \frac{17}{16} v^4 - \frac{2}{16} v^6 \right)\ln\frac{1+v}{1-v} 
- \frac{111}{24} v + \frac{119}{24} v^3 + \frac{6}{24}v^5\right],\\
\rho_-^{(0)}(s) &= {1\over 8\pi^2} v(3-2v^2)\nn\\
\rho_-^{(1)}(s) &= \frac{3}{8\pi^2}  \frac{(v^2-1)(1 - 2 v^2)}{v} \ln \frac{\mu^2}{\bar{m}^2} \nonumber \\
&+ \frac{1}{6\pi^2} \left[(3-2v^2) B(v) + \left(\frac{61}{16} + \frac{28}{16}v^2 - \frac{31}{16} v^4 - \frac{2}{16} v^6 \right)
\ln\frac{1+v}{1-v} - \frac{3}{v} + \frac{321}{24} v - \frac{241}{24} v^3 + \frac{6}{24}v^5\right],
\label{eq:PIqcd}
\end{align}
where the function $B$ is defined as 
\begin{align}
B(v) =& (1 + v^2) \left\{\frac{\pi^2}{4} + \frac{1}{2} \Li_2 \left[\left(\frac{1-v}{1+v}\right)^2\right] - 
    \frac{1}{2} \Li_2 \left(\frac{4v}{(1+v)^2}\right) - \Li_2 \left(\frac{2v}{1+v}\right) + 
    \Li_2\left(\frac{1-v}{1+v}\right)\right\} \nonumber \\
    &+ 3 v \ln\left(\frac{1-v^2}{4v}\right) - v \ln(v),
\end{align}
and we use the notation $v=\sqrt{1-\displaystyle{4m_c^2\over s}}$ and $\Li_2(x)$ is the dilogarithm function.~\footnote{For an easier comparison of the results, we should emphasize that, numerically, the above function $B(v)$ is the same as $A(u)$ function in \cite{Reinders:1981si}.}
In all of the above expressions  the quark mass, $m_c\equiv m_c^\msbar (m_c)$, and $\alpha_s = \alpha_s(\mu)$. Conversion to the pole mass can be made by using 
\bea 
\frac{M_c}{m_c(\mu)} = 1 + \left[  \frac{4}{3} + \ln\frac{\mu^2}{m_c(\mu)^2} \right] \frac{\alpha_s (\mu)}{\pi} + \; ...
\eea
We checked that our results for $\rho_{P,V}^{\rm pert}$ agree with those given in the literature, cf. eg.~\cite{Reinders:1981si}. Our expression for $\rho_+^{\rm pert}$ agrees with a similar expression in ref.~\cite{Reinders:1981si} derived by using $\bar{c} \partial_{\mu} \gamma_5 c$ instead of the tensor density $\bar{c} \sigma_{\mu\nu}c$. Expressions for $\rho_-^{\rm pert}$ are new.

\begin{figure}[htb]
\begin{center}
{\resizebox{6.9cm}{!}{\includegraphics{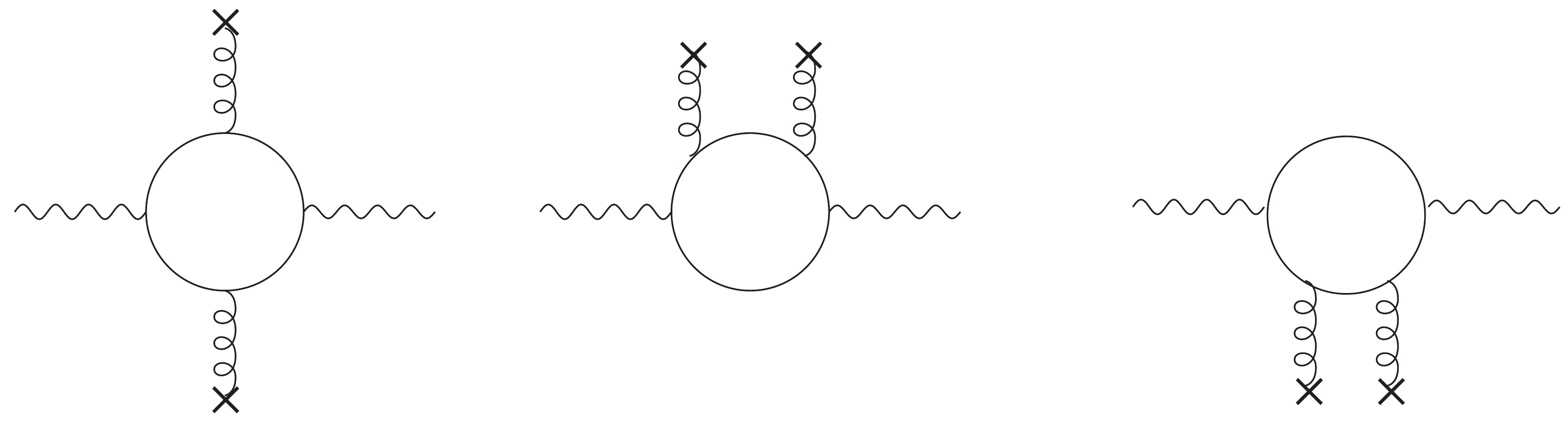}}}
\caption{\label{calc2}{\footnotesize 
Gluon condensate contribution to the spectral functions. 
 } }
\end{center}
\end{figure}
%
The nonperturbative contributions proportional to the gluon condensate are obtained by computing the diagrams shown in fig.~\ref{calc2}. In the notation  
\bea
\Pi_i^{\rm nonpert}(Q^2) \equiv \langle \frac{\alpha_s}{\pi} G^2 \rangle C_i^{\rm G}(Q^2)
\eea 
our results read as follows:~\footnote{In the literature there is some discrepancy among results for $C_i^{\rm G}(Q^2)$ related to a different 
number of subtractions. For example, in ref.~\cite{Reinders:1981si},  $C_P^{\rm G}(Q^2)$ and $C_S^{\rm G}(Q^2)$ (connected with $C_+^{\rm G}(Q^2)$ here)
 are obtained by using two-times subtracted spectral function, namely $\Pi_i(Q^2) - \Pi_i(0) - Q^2 \Pi_i(0)^{\prime}$. If we do the same here, our results would agree with 
their expressions. Similarly, in that way, $C_{A^\prime}^{\rm G}$ from ref.~\cite{Reinders:1981si} would coincide with our $C_+^{\rm G}$.}
%
\bea
C_P^{\rm G}(Q^2) =  \frac{-1}{48 Q^2} \left [ \frac{ 3 (1+ 3 v^2)(1- v^2)}{v^5} \frac{1}{2} \log \frac{1+v}{1-v}  - \frac{7v^2+3}{v^4} \right ] \,,
\eea
\bea
C_V^{\rm G}(Q^2) = \frac{1}{48 Q^4} \left [ \frac{ 3 ( 1+ v^2)(1- v^2)^2}{v^5} \frac{1}{2} \log \frac{1+v}{1-v} - \frac{ 3 v^4 - 2 v^2 + 3}{ v^4} \right ] \,,
\eea
\bea
C_+^{\rm G}(Q^2) =  
- \frac{1}{ 48 Q^4} \left [ \frac{ ( 3+ v^2)(1- v^2)}{v^3} \frac{1}{2} \log \frac{1+v}{1-v}  - \frac{3-v^2}{v^2} \right ] \,,
\eea
\bea
C_-^{\rm G}(Q^2) = \frac{1}{48 Q^4} \left [ \frac{(1-v^2)(3-7 v^2)(1 + 2 v^2)}{v^5} \frac{1}{2} \log \frac{1+v}{1-v}  - 
\frac{14 v^4 - 3 v^2 + 3 }{v^4} \right ] \,,
\eea
The above expressions are obtained by the direct calculation and agree with refs.~\cite{radyushkin,broadhurst}. The result for $C_-^{\rm G}(Q^2)$ is new.  

For calculation of the moments, the integral representation of the above formulas are particularly useful~\cite{Shifman:1978bx,nikolaev}. 
With the help of
\bea
I_N(\xi) = \int_0^1 \frac{d x}{ \left [ 1 + 4 x (1-x) \xi \right ]^N }\,,
\eea 
where $\xi = Q^2/( 4 m_c^2)$, we can express all $C^{\rm G}(Q^2)$ as 
\bea
C_P^{\rm G}(Q^2) = \frac{1}{24 Q^2} (5 + 6 I_1 -15 I_2 +4 I_3 )\,,
\eea
\bea
C_V^{\rm G}(Q^2) = \frac{1}{12 Q^4} (-1 + 3 I_2 - 2 I_3 )\,,
\eea
\bea
C_+^{\rm G}(Q^2) = \frac{1}{3 Q^2}C_S^{\rm G}(Q^2) = - \frac{1}{24 Q^4} (-1 -2 I_1 +3 I_2) \,,
\eea
\bea
C_-^{\rm G}(Q^2) =- \frac{1}{24 Q^4} (7 - 6 I_1 -5 I_2 + 4 I_3) \,.
\eea

\end{document}